\documentclass[a4paper]{article}  
\usepackage{graphicx,amsmath,amssymb,dsfont,bm}
\usepackage{color}
\usepackage{enumitem}
\usepackage{boldline}
\usepackage{geometry}
\usepackage{authblk}
\usepackage[colorlinks=true,urlcolor=black,linkcolor=blue,citecolor=blue,bookmarks=false]{hyperref}
\usepackage{listings}
\usepackage{subfig}
\usepackage[normalem]{ulem}
\usepackage{cite}
\usepackage{multirow}
\usepackage{soul}
\usepackage[title]{appendix}
\usepackage[export]{adjustbox}

\usepackage{etoolbox}
\makeatletter
\patchcmd{\l@section}
  {\hfil}
  {\leaders\hbox{\normalfont$\m@th\mkern \@dotsep mu\hbox{.}\mkern \@dotsep mu$}\hfill}
\makeatother

\title{\bf Congestions and Spectral Transition in Time-Lagged Correlations of Motorway Traffic}

\author{Gabor B. Hollbeck, Ren\'e Pilarczyk, Shanshan Wang \thanks{shanshan.wang@uni-due.de}, \\ Michael Schreckenberg, and Thomas Guhr}

\affil{\textit{Faculty of Physics, University of Duisburg--Essen, Lotharstra\ss e 1, 47048 Duisburg, Germany}}

\date{\today}
 
\begin{document}
\maketitle
\newcommand\redsout{\bgroup\markoverwith{\textcolor{red}{\rule[0.5ex]{2pt}{1pt}}}\ULon}

\noindent {\bf Abstract.} 
The congestion of a motorway section is propagated to its neighbouring  sections, leading to correlations. The resulting correlation matrix encodes the information on congestion. Here, we study symmetrized time-lagged correlations and show how their spectral properties reveal congestion durations. We carry out an empirical analysis and find a transition behavior for the dominant eigenvalue as function of the time lag. Furthermore, we set up a numerical simulation model for indicator time series of traffic phases as well as a simplified model that we treat analytically. We consider various scenarios. Our results reveal a nonlinear relation between the spectral transition and the congestion duration. In our simplified model, we derive this relation analytically.

\vspace{0.5cm}

\vspace{0.5cm}

\noindent\rule{\textwidth}{1pt}
\vspace*{-1cm}
{\setlength{\parskip}{0pt plus 1pt} \tableofcontents}
\noindent\rule{\textwidth}{1pt}

\section{Introduction}
\label{sec1}

Many factors trigger traffic congestions, such as traffic accidents, bottlenecks, commute flows, special events, bad weather etc. A report of the U. S. Federal Highway Administration reveals that the duration of weekday peak-period congestion increases from 4.5 hours per day in 1982 to 7.0 hours per day in 2003 for the largest US cities~\cite{Systematics2005}. In 2022, the drivers in London spent an average of 156 hours in traffic delays~\cite{Inrix2022}. Inrix 2022 Global Traffic Scorecard manifests that drivers in USA, UK and Germany lost, on average, dozens of hours in 2022, leading to considerable expenses in time and fuel costs~\cite{Inrix2022}. Nowadays, a wealth of data is available that awaits statistical analysis. Here we focus on the interplay between correlations and congestion durations. 

A traffic network as a whole can be viewed as a complex system partly similar to others, e.g., financial markets~\cite{Munnix2012}, human brains~\cite{Telesford2011}, power systems~\cite{Messina2009}, global climate~\cite{Cheng2014}, and so on~\cite{Podobnik2010}. A traffic network comprises many constituents, e.g., road sections coupled via the motion of vehicles~\cite{Saeedmanesh2017,Bellocchi2020,Saberi2020}, resulting in real-time traffic flows and velocities. The time series of traffic flows or velocities exhibit non-stationary characteristics and non-Markovian features at a certain time scale~\cite{Wang2020,Krause2017,Wang2023a}, distinguished by different traveling behaviors on workdays and holidays, seasonality, the presence of trucks, road construction and so on. The traffic flows also feature anti-persistent behavior, which has an impact on the congestion duration at different time scales~\cite{Krause2017}. Due to the  propagation of traffic flows~\cite{Kerner2021}, all its features generate correlations between the sections in the networks. There are spatial~\cite{Gartzke2022} and temporal~\cite{Wang2020} correlations. Spectral analyses of correlation matrices reveal the emergence of collective behavior~\cite{Wang2021} and subdominant collective behavior~\cite{Wang2022a} in large traffic networks. Statistical responses of the velocities and flows to congestions~\cite{Wang2023b,Gartzke2023} have also been studied.

The motion of vehicles among different road sections consumes time, giving rise to a time delay in the propagation of congestion effects and thus to non-Markovian statistics. Therefore, it is very interesting to explore time-lagged correlations. The time-lagged correlations~\cite{Podobnik2010} are widely used in many fields, including traffic. For motorway traffic, they have been used for the predication of traffic flows or velocities~\cite{Yang2014,Fabritiis2008}, the identification and prediction of traffic phases~\cite{Neubert1999,Pan2013}, for understanding the velocity change over space and time~\cite{Guo2020}, estimating the congestion propagation speed between neighbouring sections~\cite{Zheng2010}, studying the effects of upstream and downstream traffic on the section between them~\cite{Chandra2008,Sun2014}, spectral analyses on oscillation patterns of traffic data~\cite{Lam1970}, and correlation analyses for different motorway lanes~\cite{Knospe2002a,Knospe2002b}, etc. The above studies, however, are restricted to a pair of motorway sections. To analyze a motorway network as a whole that comprises many sections, the investigation of the entire corresponding correlation matrix is called for.

Our previous studies~\cite{Wang2020,Wang2021,Gartzke2022,Wang2022a} of large correlation matrices for traffic networks were restricted to equal-time correlations. Here, we extend this to time-lagged correlations. We dissect the empirical correlation matrices and study the statistics of the spectra. We identify spectral transitions and show how they relate to congestion durations. We perform empirical analyses and, moreover, introduce an indicator model which we analyze numerically and which we solve analytically in a simplified version. 

The paper is organized as follows. In section~\ref{sec2}, we describe our traffic data, sketch the methods and show empirical results. In section~\ref{sec3}, we set up an indicator model and perform simulations to explore the relation between congestion durations and transitions of eigenvalues as function of time lag. We then analytically solve a simplified version of the model in section~\ref{sec4}. We draw our conclusions in section~\ref{sec5}.

\section{Spectral transitions in time-lagged correlations in the presence of congestions}
\label{sec2}

We describe the used traffic data in section~\ref{sec21}. We introduce symmetrized original and extended correlation matrices in section~\ref{sec22}. Furthermore, we display the evolution behavior of their eigenvalues for two motorway networks in section~\ref{sec23}.

\subsection{Data description}
\label{sec21}

We use the traffic data collected by inductive loop detectors on 679 motorway sections in the Rhine-Ruhr metropolitan region and on 42 motorway sections around Breitscheid, all in North Rhine-Westphalia (NRW), Germany, see figure~\ref{fig1}. The data with a resolution of one minute contains the information on time, traffic flow (i.e., the number of vehicles passing a detector per unit time) and velocity for every lane of each section. 

As described in references~\cite{Wang2021,Wang2020, Wang2022a}, a motorway network exhibits different behavior on workdays and on holidays, resulting in different correlation structures between motorway sections. Here, we only use the traffic data on workdays, and focus on the afternoon rush hours (15:00--20:00), when the commute flows are more likely to produce traffic congestion. The fundamental diagram in traffic theory~\cite{Kerner2012} implies that a specific traffic flow value may correspond to free as well as to congested flow, while a specific velocity value must correspond to either free or congested flow. In the latter case, it is easier to distinguish different traffic phases. Therefore, we use velocity time series to work out the empirical correlations between motorway sections. 

Missing values in a data matrix will influence the spectral decomposition of the corresponding correlation matrix. To avoid this, we fill the missing values of each time series by linear interpolation of the nearest non-zero values. At the beginning or at the end of a time series, we replace the missing values with the subsequent or previous values, respectively.

\begin{figure}[tbp]
\centering
\includegraphics[width=0.46\linewidth]{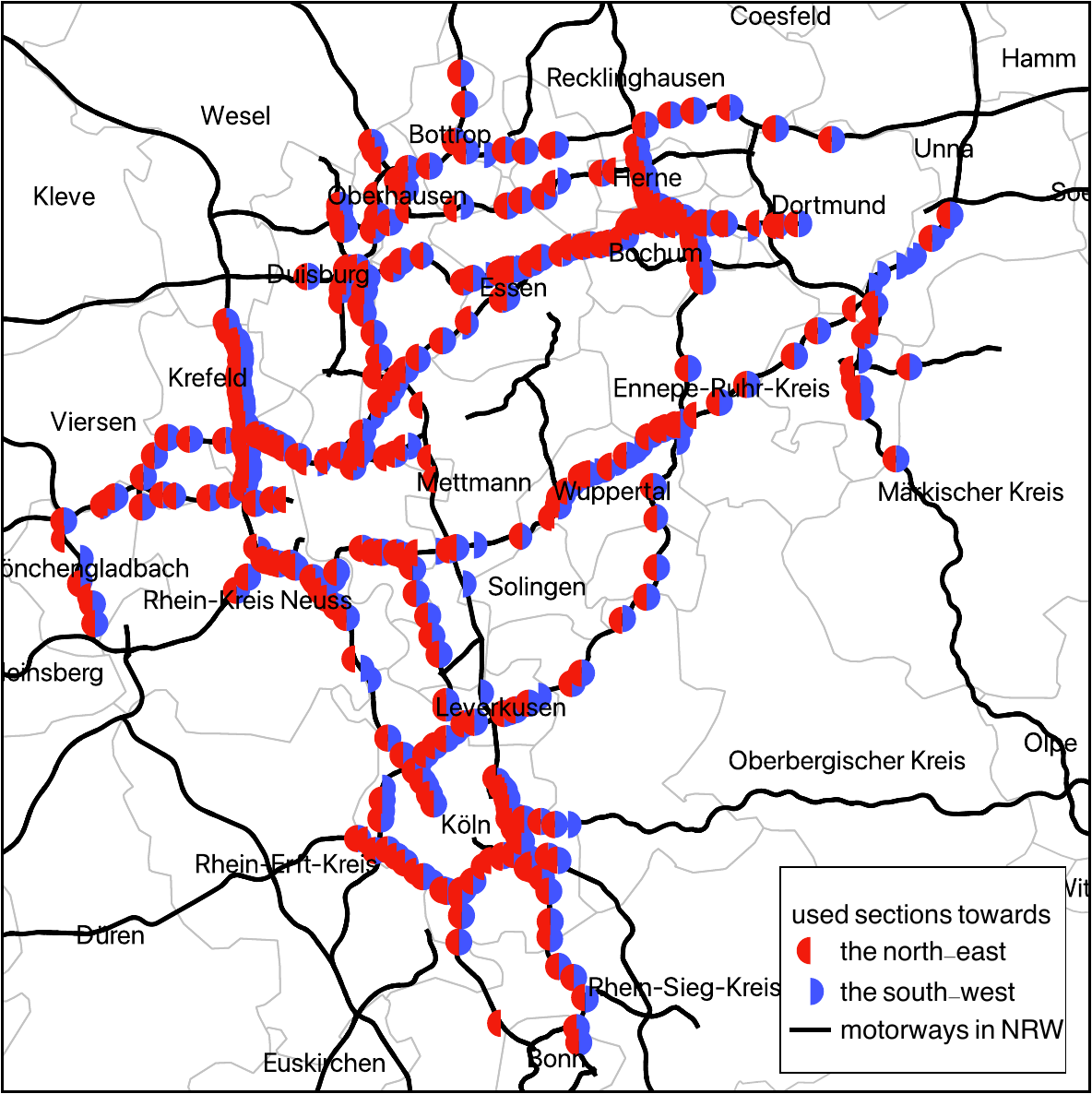}
\includegraphics[width=0.46\linewidth]{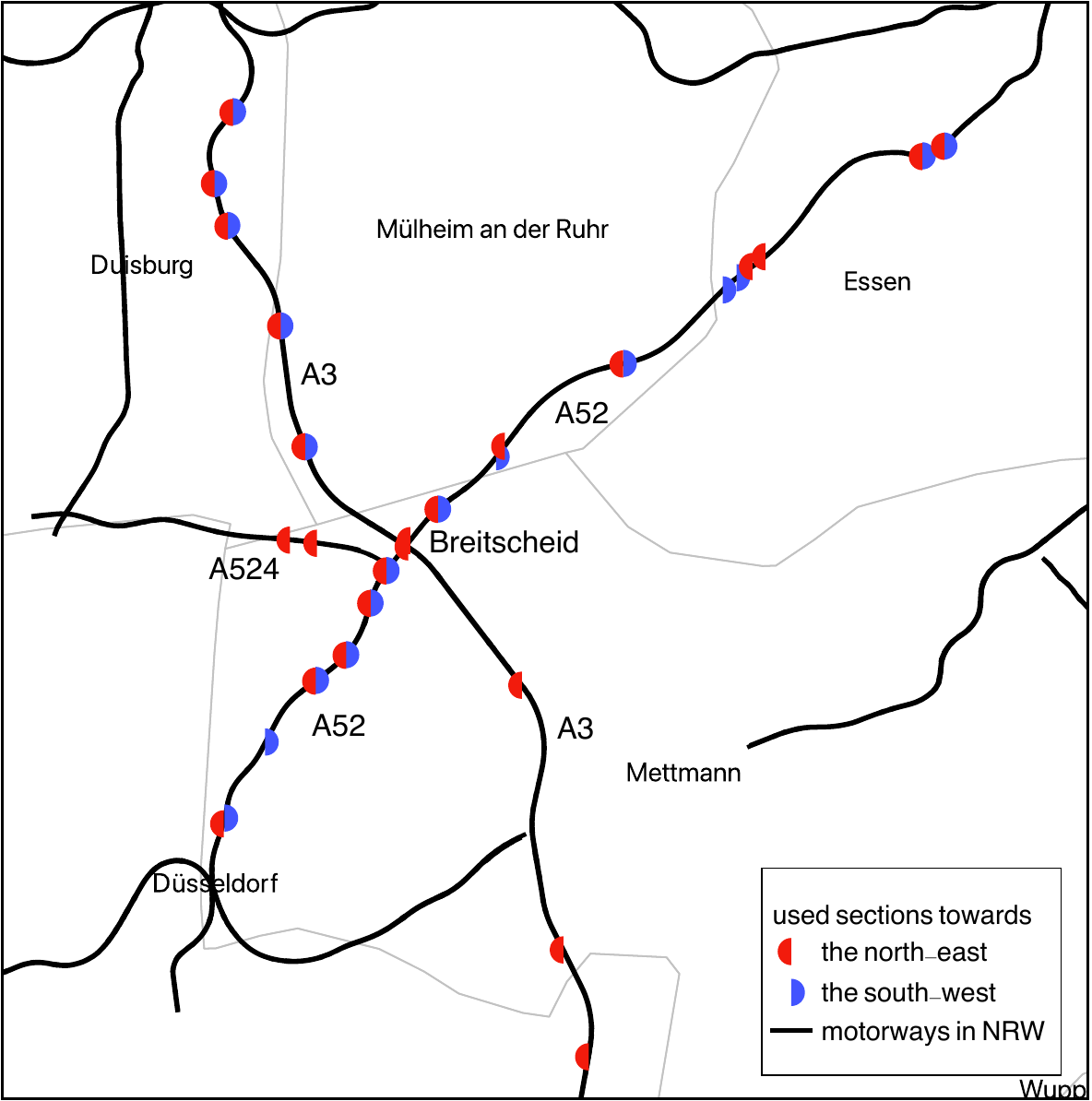}
\caption{Geographic positions of detectors in the Rhine-Ruhr metropolitan region and in the Breitscheid region in NRW, Germany. The base map and data are from OpenStreetMap $\copyright$ OpenStreetMap contributors, licensed under the Open Database License (ODbL) \cite{osmcopyright,osm}. The maps are generated with the geographic information system software QGIS 3.4 \cite{qgis}. }
\label{fig1}
\end{figure}

\subsection{Time-lagged correlation matrices}
\label{sec22}

A time series of velocities in section $k$ with length $T$ is denoted by $v_k(t)$. There are $K$ such time series $v_k(t)$, $k=1,\cdots,K$, which we order in a rectangular $K\times T$ data matrix
\begin{equation}
G^\mathrm{(o)}=\left[ \begin{array}{ccc}
v_1(1) 	& \hdots 	& v_1(T) 		\\
\vdots	&  \ddots	& \vdots			\\
v_K(1)	& \hdots  	& v_K(T) 
\end{array}
\right ] .
\label{eq2.1.1}
\end{equation}
The time series of section $k$ is the $k$-th row of $G^\mathrm{(o)}$, where the superscript $\mathrm{(o)}$, anticipating later usage, refers to the original data. Here $K=679$ for the Rhine-Ruhr metropolitan region and $K=42$ for the region around Breitscheid. We introduce two $T\times T$ projection matrices 
\begin{eqnarray}
P_{T,\tau}=
\left[\begin{array}{cc}
\mathds{1}_{(T-\tau)\times(T-\tau)} & \mathcal{O}_{\tau\times(T-\tau)}\\
\mathcal{O}_{(T-\tau)\times \tau} & \mathcal{O}_{\tau\times\tau}\\
\end{array}\right]
\quad
\mathrm{and}
\quad
Q_{T,\tau}=
\left[\begin{array}{cc}
\mathcal{O}_{\tau\times\tau}&\mathcal{O}_{(T-\tau)\times \tau} \\
 \mathcal{O}_{\tau\times(T-\tau)}& \mathds{1}_{(T-\tau)\times(T-\tau)} \\
\end{array}\right]
\label{eq2.1.4}
\end{eqnarray}
where $\mathds{1}_{(T-\tau)\times(T-\tau)}$ is a $(T-\tau)\times(T-\tau)$ identity matrix, and $\mathcal{O}_{\tau\times(T-\tau)}$, $\mathcal{O}_{(T-\tau)\times \tau}$ and $\mathcal{O}_{\tau\times\tau}$ are $\tau\times(T-\tau)$, $(T-\tau)\times \tau$ and $\tau\times\tau$ zero matrices, respectively. With equation~\eqref{eq2.1.4}, the time series from time $1$ to $T-\tau$ in $G^\mathrm{(o)}$ can be written as the $2\times 2$ matrix $G^\mathrm{(o)}P_{T,\tau}$ and from $1+\tau$ to $T$ in $G^\mathrm{(o)}$ as the $2\times 2$ matrix $G^\mathrm{(o)}Q_{T,\tau}$.

For a time series from time $t_a$ to time $t_b$, mean value $\mu_{k}$ and standard deviation $\sigma_{k}$ read
\begin{equation}
\mu_{k}=\frac{1}{t_b-t_a+1} \sum_{t=t_a}^{t_b} v_{k}(t)
\quad
\mathrm{and} 
\quad
\sigma_{k}=\sqrt{\frac{1}{t_b-t_a+1} \sum_{t=t_a}^{t_b}\left(v_{k}(t)-\mu_{k}\right)^{2}} \ ,
\label{eq2.1.2}
\end{equation}
respectively. 
We normalize each time series to zero mean and unit standard deviation by 
\begin{equation}
M_{k}^\mathrm{(o)}(t)=\frac{v_{k}(t)-\mu_{k}}{\sigma_{k}}  \ ,
\label{eq2.1.3}
\end{equation}
and introduce $M^\mathrm{(o)}_P$ and $M^\mathrm{(o)}_Q$ as the $K\times (T-\tau)$ normalized data matrices of $G^\mathrm{(o)}P_{T,\tau}$ and $G^\mathrm{(o)}Q_{T,\tau}$, respectively. This yields a $K  \times K$ time-lagged correlation matrix with time lag $\tau$,
\begin{equation}
C^\mathrm{(o)}(\tau)= \dfrac{1}{T-\tau} M^\mathrm{(o)}_P\Big(M^\mathrm{(o)}_Q\Big)^{\dagger}  \ ,
\label{eq2.1.5}
\end{equation}
where $\dagger$ indicates the transpose, such that
\begin{equation}
C_{kl}^\mathrm{(o)}(\tau)= \dfrac{1}{T-\tau} \sum_{t=1}^{T-\tau} M_k^\mathrm{(o)}(t)M_l^\mathrm{(o)} (t+\tau) \  
\label{eq2.1.6}
\end{equation}
is the time-lagged correlation coefficient between sections $k$ and $l$. We notice that the sum runs from $t=1$ to $t=T-\tau$. Consider the motorway sections in a data matrix labeled by numbers in an ascending order. For $k<l$, $C_{kl}^{(o)}(\tau)$ in equation~\eqref{eq2.1.6} measures the time-lagged correlation between sections $k$ and $l$, where the velocity in section $l$ lags behind the one in section $k$ at time $t$. Considering the velocity in section $l$ at an earlier time $t$, the time lag $\tau$ requires the corresponding velocity on section $k$ occurring also at an earlier time $t-\tau$. Therefore we can reformulate $C_{kl}^{(o)}(\tau)$ as 
\begin{equation}
C_{kl}^\mathrm{(o)}(\tau)= \dfrac{1}{T-\tau} \sum_{t=\tau+1}^{T} M_k^\mathrm{(o)}(t-\tau)M_l^\mathrm{(o)} (t) \ ,
\label{eq2.1.6.1}
\end{equation}
where now the sum runs from $t=\tau+1$ to $t=T$. Equation~\eqref{eq2.1.6.1} can be regarded as the time-led correlation coefficient between sections $k$ and $l$ with reference time $t$. For $k=l$, $C_{kl}^\mathrm{(o)}(\tau)$ either as in equation~\eqref{eq2.1.6} or as in equation~\eqref{eq2.1.6.1} gives the autocorrelation of section $k$ or $l$. When $\tau=0$, the correlation matrix $C^\mathrm{(o)}(0)$ reduces to the standard equal-time Pearson correlation matrix which is symmetric and positive semidefinite with real and positive eigenvalues. When $\tau>0$, $C^\mathrm{(o)}(\tau)$ is asymmetric, leading to complex eigenvalues. The symmetrized time-lagged correlation matrix
\begin{equation}
\bar{C}^\mathrm{(o)}(\tau)= \frac{1}{2} \Big(C ^\mathrm{(o)}(\tau) +\big( C^\mathrm{(o)}(\tau)\big)^{\dagger} \Big) \ ,
\label{eq2.1.7}
\end{equation}
however has real eigenvalues which can be negative. The elements 
\begin{equation}
\bar{C}_{kl}^\mathrm{(o)}(\tau)= \frac{1}{2} \Big(C_{kl} ^\mathrm{(o)}(\tau) +C_{lk}^\mathrm{(o)}(\tau) \Big) \ ,
\label{eq2.1.7.1}
\end{equation}
of the matrix $\bar{C}^\mathrm{(o)}(\tau)$ is the average of the time-lagged correlation and the time-lead correlation, and the scenarios $k<l$ and $k>l$ are not distinguished.

\begin{figure}[tbp]
\centering
\includegraphics[width=0.8\linewidth]{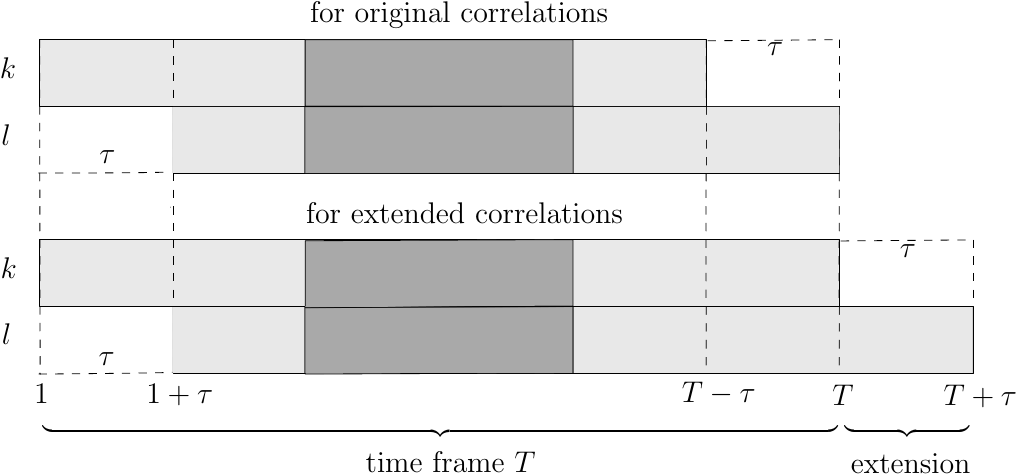}
\caption{Two time series for the calculation of original and extended correlations with time lag $\tau$. Light and dark grey regions represent the values for free and congested phases, respectively. }
\label{fig2}
\end{figure}

Equation~\eqref{eq2.1.5} is widely used for calculating correlation matrix with a time lag. A drawback of this equation is that the length $T-\tau$ of time series for calculation becomes ever shorter with increasing $\tau$. This might give rise to poor statistics for very large $\tau$. We thus make the length of the time series used independent of $\tau$ by extending the interval between $t_a=1+\tau$ and $t_b=T$ to the interval between $t_a=1+\tau$ and $t_b=T+\tau$ for the lagged time series. Thus, the data outside the initial time series between $1$ and $T$ are included, as sketched in figure~\ref{fig2}. Here $G^\mathrm{(e)}$ is a $K \times (T+\tau)$ extended data matrix, where superscript $\mathrm{(e)}$, anticipating the later usage, refers to the extended data. Accordingly, $P_{T+\tau,\tau}$ and $Q_{T+\tau,\tau}$ are two $(T+\tau)\times (T+\tau)$ projecting matrices. We use $M^\mathrm{(e)}_P$ and $M^\mathrm{(e)}_Q$ to represent the $K\times T$ normalized data matrices of $G^\mathrm{(e)}P_{T+\tau,\tau}$ and $G^\mathrm{(e)}Q_{T+\tau,\tau}$. In contrast to the lagged time series, the leading time series is fixed from $t_a=1$ and $t_b=T$ for any $\tau$. We refer to the correlations between the leading and lagged time series as extended correlations. The extended time-lagged correlation matrix is
\begin{equation}
C^\mathrm{(e)} (\tau)= \dfrac{1}{T}M^\mathrm{(e)}_P\Big(M^\mathrm{(e)}_Q\Big)^{\dagger} ,
\label{eq2.1.8}
\end{equation}
with the elements given by
\begin{equation} 
C_{kl}^\mathrm{(e)}(\tau)= \dfrac{1}{T} \sum_{t=1}^{T} M_k^\mathrm{(e)}(t)M_l^\mathrm{(e)} (t+\tau) \ , 
\label{eq2.1.9}
\end{equation} 
or 
\begin{equation}
 C_{kl}^\mathrm{(e)}(\tau)= \dfrac{1}{T} \sum_{t=\tau+1}^{T+\tau} M_k^\mathrm{(e)}(t-\tau)M_l^\mathrm{(e)} (t) \ .
\label{eq2.1.10}
\end{equation}
As in equation~\eqref{eq2.1.7}, the symmetrization of this correlation matrix results in $\bar{C}^\mathrm{(e)}(\tau)$, of which each element is the average of the time-lagged correlation and the time-led correlation.

\subsection{Spectral properties as functions of the time lag}
\label{sec23}

To study the spectra, we diagonalize the symmetrized, hence real-symmetric, matrices,
\begin{equation}
\bar{C}^\mathrm{(x)}(\tau)=U \Lambda U^{\dagger} \ , \qquad \mathrm{x}=\mathrm{e,o}\ ,
\label{2.3.1}
\end{equation}   
where the diagonal matrix $\Lambda=\mathrm{diag}(\Lambda_{1},\Lambda_{2},\cdots, \Lambda_{K})$ contains the real eigenvalues $\Lambda_{k}$ of $\bar{C}^\mathrm{(x)}(\tau)$. All eigenvalues are ranked in an ascending order, such that $\Lambda_{1}$ and $\Lambda_{K}$ are the smallest and the largest eigenvalues, respectively. The $k$-th column  of the $K  \times K$ orthogonal matrix $U$ is the eigenvector $U_{k}$ to the eigenvalue $\Lambda_k$. The non-symmetrized matrices $C^\mathrm{(x)}(\tau)$ are real, but not symmetric for $\tau>0$. Thus they can be decomposed as $C^\mathrm{(x)}(\tau)=V_1SV_2^{\dagger}$, where the diagonal matrix $S$ contains the singular values. The orthogonal matrices $V_1$ and $V_2$ are different from each other. It may happen that $C^\mathrm{(x)}(\tau)$ for $\tau>0$ can be diagonalized in the form $C^\mathrm{(x)}(\tau)=u\lambda u^{\dagger}$ where $u$ is real and orthogonal and the diagonal matrix $\lambda$ contains complex eigenvalues. Only in this case, we have 
\begin{equation}
\bar{C}^{\mathrm{(x)}}=\frac{1}{2}(u\lambda u^{\dagger}+u \lambda^\dagger u^{\dagger})=u\frac{\lambda+\lambda^\dagger}{2}u^{\dagger}=u(\mathrm{Re}\lambda)u^{\dagger}  \ ,
\end{equation}
implying that the eigenvalues of the symmetrized time-lagged correlation matrices are the real values of those of the original time-lagged correlation matrices $C^\mathrm{(x)}(\tau)$. In general, however, such the diagonalization $C^\mathrm{(x)}(\tau)=v\lambda v^{\dagger}$ for $\tau>0$ will involve unitary matrices $v$, and the real parts of the eigenvalues will not coincide with those of the symmetrized $\bar{C}^\mathrm{(x)}(\tau)$.

\begin{figure}[tbp]
\centering
\includegraphics[width=\linewidth]{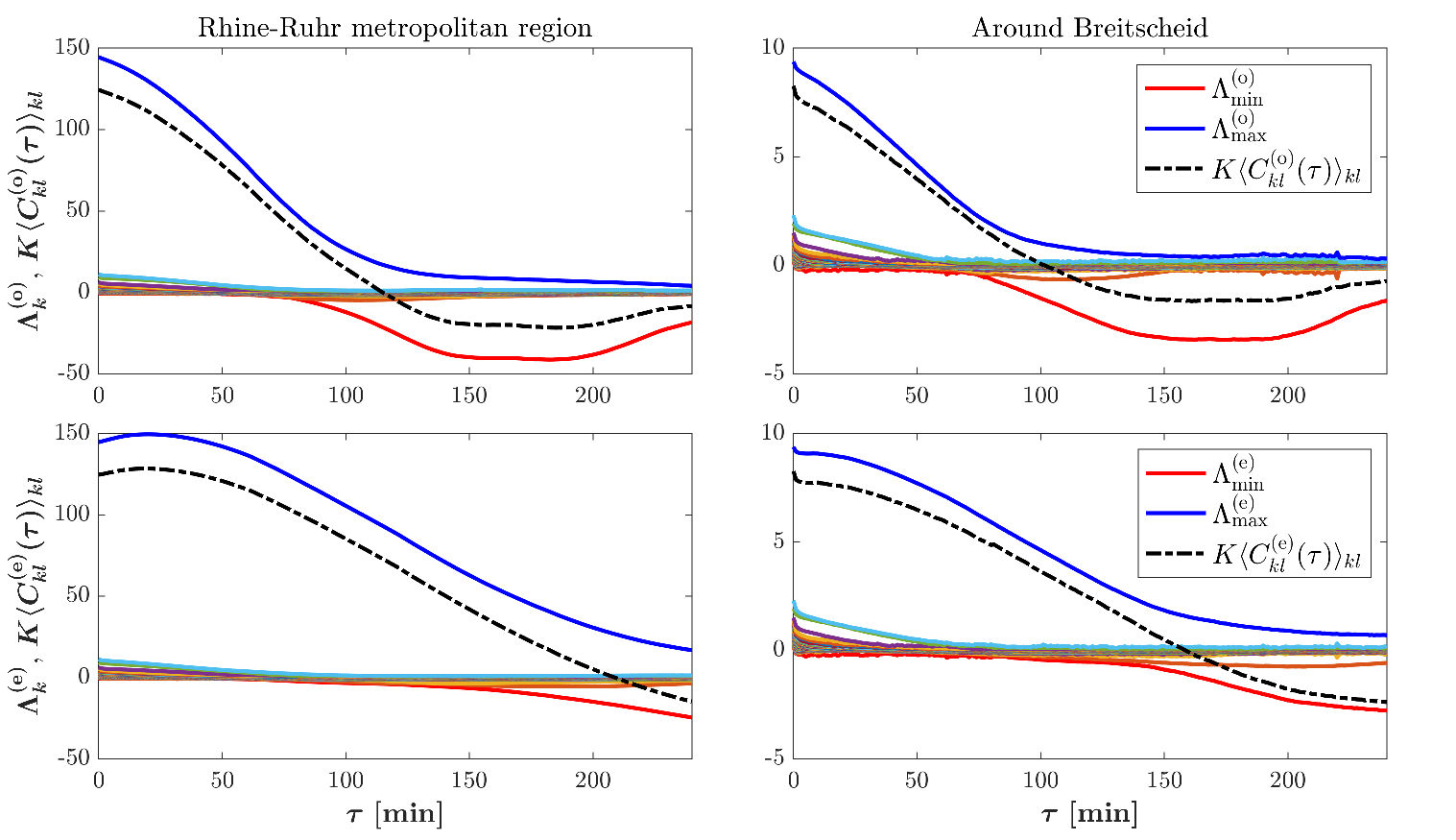}
\caption{Eigenvalues from time-lagged correlation matrices $\bar{C}^\mathrm{(o)}(\tau)$ and $\bar{C}^\mathrm{(e)}(\tau)$ and $K$-fold average correlations $K\langle C_{kl}^\mathrm{(o)}(\tau)\rangle_{kl}$ and $K\langle C_{kl}^\mathrm{(e)} (\tau)\rangle_{kl}$, versus time lag $\tau$ for empirical data of the Rhine-Ruhr metropolitan region and around Breitscheid in Germany.}
\label{fig3}
\end{figure}

We work out $C^\mathrm{(o)} (\tau)$ and $C^\mathrm{(e)} (\tau)$ with empirical data of velocities during the time period from 15:00 to 20:00, i.e., $T=300$ min, for each workday. The largest time lag $\tau$ is restricted to 240 min. We further average $C^\mathrm{(o)} (\tau)$ and $C^\mathrm{(e)}(\tau)$ over 64 discontinuous workdays to remove the effect of noise and then symmetrize them for spectral decomposition. In figure~\ref{fig3}, we show the eigenvalues of $\bar{C}^\mathrm{(x)}(\tau)$, x=e or o, versus the time lag for empirical data of the Rhine-Ruhr metropolitan region as well as around the motorway intersection near Breitscheid. The largest and the smallest eigenvalues are separated from other eigenvalues that are close to zero. In contrast to the case of a correlation matrix without time lags, the smallest eigenvalue becomes negative and gradually prevails over the largest eigenvalue as the time lag becomes larger. Put differently, with $\tau$ increasing, the dominant eigenvalue shifts from the largest eigenvalue to the smallest eigenvalue. We define the spectral transition as a change of dominant eigenvalues from positive values to negative values. The transition point corresponds to the time lag at which the sign changes. Such spectral transition happens for both motorway networks. The previous study~\cite{Wang2021} on the correlations in traffic without time lags reveals that the dominant eigenvalue, i.e. the largest positive eigenvalue, indicates the collective behavior of motorway sections. Here the transition of the dominant eigenvalue from positive values to negative values might hint a potential transition in traffic behavior.

For a correlation matrix without time lags in financial markets, the largest eigenvalue is in proportion to the average of all elements in the correlation matrix~\cite{Stepanov2015,Plerou2002,Song2011,Pharasi2019}. For correlation matrices with time lags in traffic, the average correlations $\langle C_{kl}^\mathrm{(o)}(\tau)\rangle_{kl}$ and $\langle C_{kl}^\mathrm{(e)}(\tau)\rangle_{kl}$ are not exactly proportional to the largest eigenvalues of $\bar{C}^\mathrm{(o)}(\tau)$ and $\bar{C}^\mathrm{(e)}(\tau)$, but shift from being nearly proportional to the largest eigenvalues to being nearly proportional to the smallest eigenvalues, as also shown in figure~\ref{fig3}. Once more, a transition from positive values to negative values shows up in the evolution of either $K\langle C_{kl}^\mathrm{(o)}(\tau)\rangle_{kl}$ or $K\langle C_{kl}^\mathrm{(e)}(\tau)\rangle_{kl}$. The transition occurs when the average correlation vanishes, i.e., $K\langle C_{kl}^\mathrm{(o)}(\tau)\rangle_{kl}\approx 0$ or $K\langle C_{kl}^\mathrm{(e)}(\tau)\rangle_{kl}\approx 0$. It is worth mentioning that the transition in the original cases is earlier than in the extended ones, see figure~\ref{fig3}. This difference may be traced back to the definition of extended correlations: the lagged time series includes ever more data outside the length of time series $T$ with $\tau$ increasing, effectively lowering the extended correlations than the original correlations at large $\tau$.

\section{Indicator correlations and congestion duration}
\label{sec3}

The empirical results manifest a transition in the time evolution of eigenvalues. Is this transition related to traffic congestion? If yes, what kind of relation? One of the quantities to be measured is the congestion duration, a time period during which the velocity is below a threshold, e.g. 60 km/h. Thus, we can answer the questions just posed by replacing the time series of velocities with those of indicators for free or congested traffic, i.e. of traffic phases. This facilitates the construction of a simple model that we numerically analyze in five scenarios with different congestion durations and different shapes of congestion propagation in section~\ref{sec31}. We observe transition points in the evolution of eigenvalues and explore their relation to congestion duration in section~\ref{sec32}.

\subsection{Spectral transitions related to congestion durations}
\label{sec31}

\begin{figure}[tbp]
\centering
\includegraphics[width=1\linewidth]{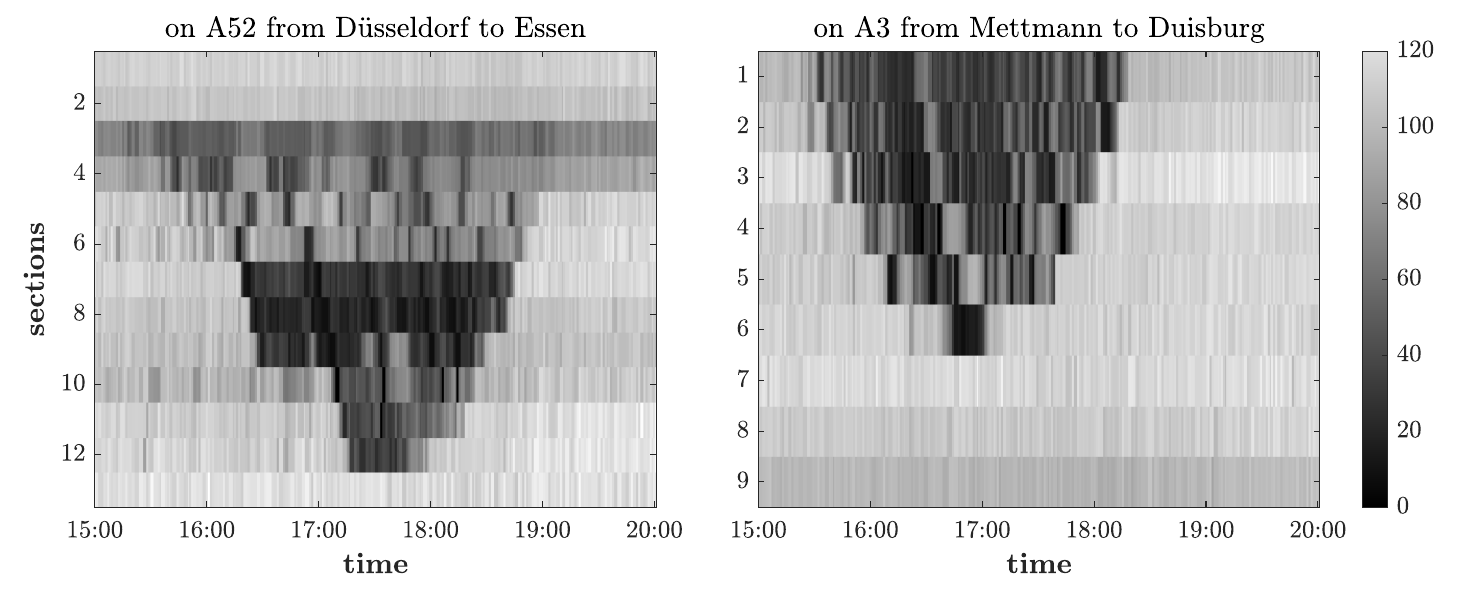}
\caption{Data matrices of velocities in km/hour on sections of motorway A52 (left) and on sections of motorway A3 (right), respectively. The direction of traffic flow on motorway A52 is from D\"usseldorf to Essen, corresponding to section 13 to section 1 in the left data matrix, while the direction of congestion propagation shown in the left data matrix is from section 1 to section 13. The direction of traffic flow on motorway A3 is from Mettmann to Duisburg, corresponding to section 9 to section 1 in the right data matrix, while the direction of congestion propagation in the right data matrix is from section 1 to section 9.}
\label{fig4}
\end{figure}

Figure~\ref{fig4} displays two empirical data matrices of velocities during afternoon rush hours on Jan. 18, 2017. The data matrices visualize the backward propagation of congestion, where the dark grey areas stand for low velocities, implying congested traffic phase, while the light grey areas for high velocities, implying free traffic phase. Both shapes of congestion propagation are similar to a triangle or a trapezium. To analyze the influences of congestion duration on the transition in the evolution of eigenvalues, we conduct five scenarios. We define an indicator of traffic phases on section $k$ at time $t$ as 
\begin{equation}
I_k(t)=\left\{ \begin{array}{ll}
	1,   &\text{for~free~phase} \ , \\
	0,  & \text{for~congested~phase}	\ ,
\end{array} \right.
\label{eq3.1.1}
\end{equation} 
which entries into a $K\times T$ data matrix $I$, simulated as follows. The time series in the data matrix $I$ are sorted in a descending oder of congestion duration of sections so that the time series with the longest congestion duration $D_\mathrm{max}$ is the top row of $I$, while the time series with the shortest duration $D_\mathrm{min}$ ($D_\mathrm{min}>0$) is the bottom row of $I$. The congestion starts first at the top section and propagates backward to the bottom section in $I$ with gradually reduced durations from $D_\mathrm{max}$ to $D_\mathrm{min}$. We also assume the time of backward propagation from one section to its nearest neighbouring section is one minute and the congestion duration is symmetrically located in the center of a time series with length $T$, thereby forming the shape of a inverted pyramid or trapezoid as $D_\mathrm{max}>D_\mathrm{min}$. Given $D_\mathrm{max}$ for the top time series and $D_\mathrm{min}$ for the bottom time series, we can determine the number of sections $K$ by
 \begin{equation}
 K=\frac{D_\mathrm{max}}{2}-\frac{D_\mathrm{min}}{2}+1 \ .
 \label{eq3.1.2} 
 \end{equation}
With a given $T$ and an obtained $K$, we can set up a $K\times T$ data matrix $I$ with binary values. Each row of the data matrix contains an indicator time series of a section, where zero is given in congestion duration and one is outside the congestion duration. Fixing the the length of time series $T=300$ min, we thus simulate five data matrices, as shown in figure~\ref{fig5}, with different parameters $D_\mathrm{max}$ and $D_\mathrm{min}$, listed in Table~\ref{tab1}. 

\begin{figure}[tbp]
\begin{center}
\includegraphics[width=\linewidth]{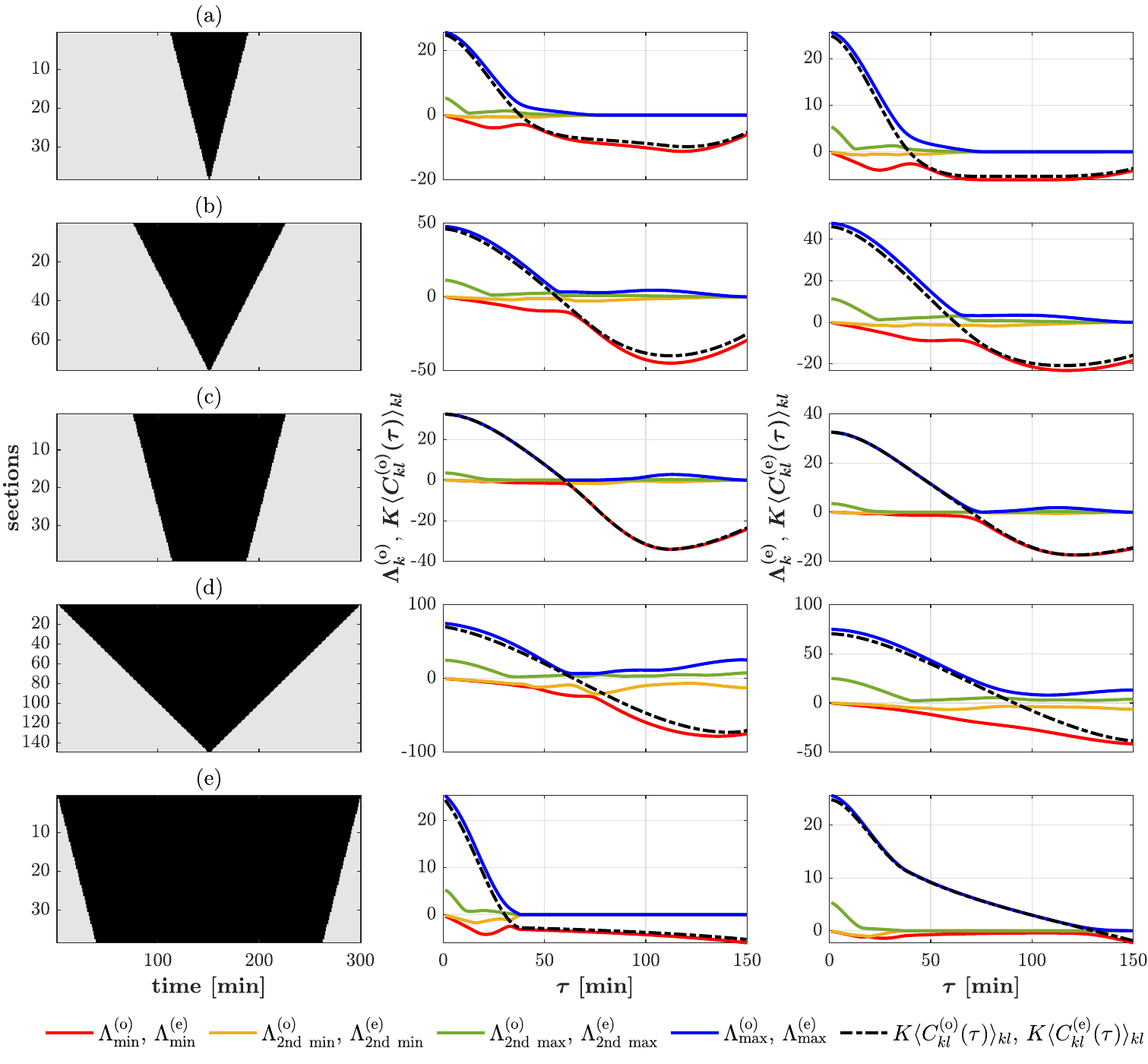}
\caption{ First column: five $K  \times T $ data matrices of indicators with symmetrical congestion and fixed length of time series $T =300$ min. Light gray color in the data matrix represents $I_{k}(t)=1$ for free phases, and black color represents $I_{k}(t)=0$ for congested phases. Second column: eigenvalues of $\bar{C}^\mathrm{(o)}(\tau)$ and $K $-fold average correlations $K\langle C_{kl}^\mathrm{(o)}(\tau)\rangle_{kl}$ versus time lag $\tau$. Third column: eigenvalues of $\bar{C}^\mathrm{(e)}(\tau)$ and $K $-fold average correlations $K\langle C_{kl}^\mathrm{(e)}(\tau)\rangle_{kl}$ versus time lag $\tau$.  The parameters of each data matrix and the transition points are listed in Table~\ref{tab1}.}
\label{fig5}
\end{center}
\end{figure}

\begin{table}[tbp]
\caption{Simulation parameters and resulting transition points for five scenarios with total length of time series $T=300$ min. Here $D_\mathrm{max}$ and $D_\mathrm{min}$ are the longest and the shortest durations, respectively, in a simulated data matrix $I$. Number of sections $K$ determined by equation~\eqref{eq3.1.2}. Transition points $\tau^\mathrm{(o)*}$ and $\tau^\mathrm{(e)*}$ of the dominant eigenvalues for original and extended correlations, respectively. }
\begin{center}
\begin{tabular*}{\textwidth}{c@{\extracolsep{\fill}}rclcccc}
\hlineB{2}
scenario & $K$ & $D_\mathrm{max}$ & $D_\mathrm{min}$ & $D_\mathrm{max}-D_\mathrm{min}$ & $\tau^\mathrm{(o)*}$& $\tau^\mathrm{(e)*}$  & $\tau^\mathrm{(e)*}-\tau^\mathrm{(o)*}$\\
 & & [min] &[min] &[min] &[min] &[min] & [min]  \\[0.5em]
\hline
(a) &38 & $~76 \approx T/4$ & ~~~2 & $~74 \approx T/4$ &37.5 &39.5 & 2 \\
 (b) &75& $150= T/2$	&~~~2 &$148 \approx T/2$ &56.5 &62.5 & 6\\
(c) &39 &$150= T/2$	& $~74\approx T/4$ &$~76 \approx T/4$ &59.5 &69.5 & 10\\ 
(d) &149& $298\approx T~~~$	& ~~~2 &$296 \approx T~~~$ &64.5&91.5 & 27\\	
(e)& 38& $298 \approx T~~~$  &$224 \approx 3T/4$ &$~74 \approx T/4$ &30.5&130.5 & 100\\

\hlineB{2}
\end{tabular*}
\end{center}
\label{tab1}
\end{table}%

With each of the five simulated data matrices, we work out the two kinds of correlation matrices $C^\mathrm{(x)}(\tau)$, x=e or o, in terms of different time lags. For extended correlations, the extended part of time series are from free phases, i.e., $I_k(t)=1$, as the congestion found during afternoon rush hours has been resolved during the long night time. Diagonalizing each correlation matrix, we obtain the eigenvalues versus time lags visualized in figure~\ref{fig5}. For the same longest congestion duration $D_{\mathrm{max}}$, e.g., $T/2$ for scenarios (b) and (c) and $T$ for scenarios (d) and (e), the different shortest durations $D_{\mathrm{min}}$ lead to different congestion shapes and distinct differences $D_\mathrm{max}-D_\mathrm{min}$. These have an influence on the remarkable transitions of the largest and the smallest eigenvalues, leading to a time gap between them. They also influence the point when the average correlation vanishes, i.e., 
\begin{equation}
K\langle C_{kl}^\mathrm{(x)}(\tau)\rangle_{kl}\approx 0 \ . 
\end{equation}
When $D_\mathrm{max}=T/2$ in scenarios (b) and (c), the remarkable transitions in the largest and the smallest eigenvalues are very close to the point at which the average correlation vanishes. Most interestingly, for scenario (c) which contains neither the smallest $D_\mathrm{min}=2$ min (e.g., in scenarios (a), (b) and (d)) nor the largest $D_\mathrm{max}=T$ (e.g., in scenarios (d) and (e)), the remarkable transitions in the largest and the smallest eigenvalues almost overlap with the vanishing of the average correlations at about 60 min. Increasing the duration to around $T$ in scenarios (d) and (e), the transitions in the largest and the smallest eigenvalues are more mild in extended correlations than in original correlations, leaving a wide gap between the two eigenvalues in extended correlations. 

As those results show, we may utilize the vanishing of the average correlations as a substitute for the transition in the dominant eigenvalues. This substitution helps to quantify the difference between transition points  $\tau^{\mathrm{(o)}*}$ and $\tau^{\mathrm{(e)}*}$ of dominant eigenvalues, which are obtained by minimizing the absolute value of the average correlation to make the average correlation as close to zero as possible, 
\begin{equation}
\tau^\mathrm{(x)*}=\mathrm{arg~min} \left |K\langle C_{kl}^\mathrm{(x)}(\tau)\rangle_{kl}\right| \ , \qquad \mathrm{x}=\mathrm{e,o}\ ,
\end{equation}
as the average correlation typically does not become exactly zero. This difference is large when $D_\mathrm{max}=T$, but relatively small when $D_\mathrm{max}\leq T/2$, as listed in table~\ref{tab1}. We further corroborate these observations by working out the transition points $\tau^\mathrm{(x)*}$ as function of the average congestion duration and the corresponding standard deviation. The average and the variance of congestion durations are given by
\begin{equation}
\langle D_k\rangle =\frac{1}{K}\sum_{k=1}^{K}D_k \ 
\quad
\mathrm{and} 
\quad
\mathrm{var}(D_k) =\frac{1}{K}\sum_{k=1}^{K}D_k^2-\left(\frac{1}{K}\sum_{k=1}^{K}D_k\right)^2\ ,
\end{equation} 
respectively. We use the standard deviation $\mathrm{std}(D_k)=\sqrt{\mathrm{var}(D_k)}$.

We let $D_\mathrm{min}$ run from 2 to $L-2$ and $D_\mathrm{max}$ from $D_\mathrm{min}+2$ to $L$ with a duration step of 2 min. This generates different combinations of $D_\mathrm{min}$ and $D_\mathrm{max}$. Each combination of $D_\mathrm{min}$ and $D_\mathrm{max}$ yields a data matrix, which results in an average congestion duration $\langle D_k\rangle$, a standard deviation of congestion durations $\mathrm{std}(D_k)$, and a transition point $\tau^\mathrm{(x)*}$ with x=o or e. Figure~\ref{fig6} shows a nonlinear dependency of $\tau^\mathrm{(x)*}$ on $\langle D_k\rangle$ and $\mathrm{std}(D_k)$. For original and extended correlations, with $L=T=300$ min, the dependencies are very similar when $\langle D_k\rangle \leq T/2$ but rather different when $\langle D_k\rangle>T/2$. In the latter case, the extended correlation might be distorted by appending the indicators of free traffic phases to the lagged indicator time series, or the vanishing of the extended correlations might not correctly reflect the transition point of the dominant eigenvalue. We find large standard deviations around $\langle D_k\rangle=L/2$. The longer the $L$ considered, the larger the differences $D_\mathrm{max}-D_\mathrm{min}$ and the standard deviations. For $L=T/2=150$ min, the original and the extended correlations show similar dependencies with relatively low standard deviations. Thus, we restrict our study to the scenarios with $D_\mathrm{max}\leq T/2$ in the sequel.

\begin{figure}[tbp]
\begin{center}
\includegraphics[width=\textwidth]{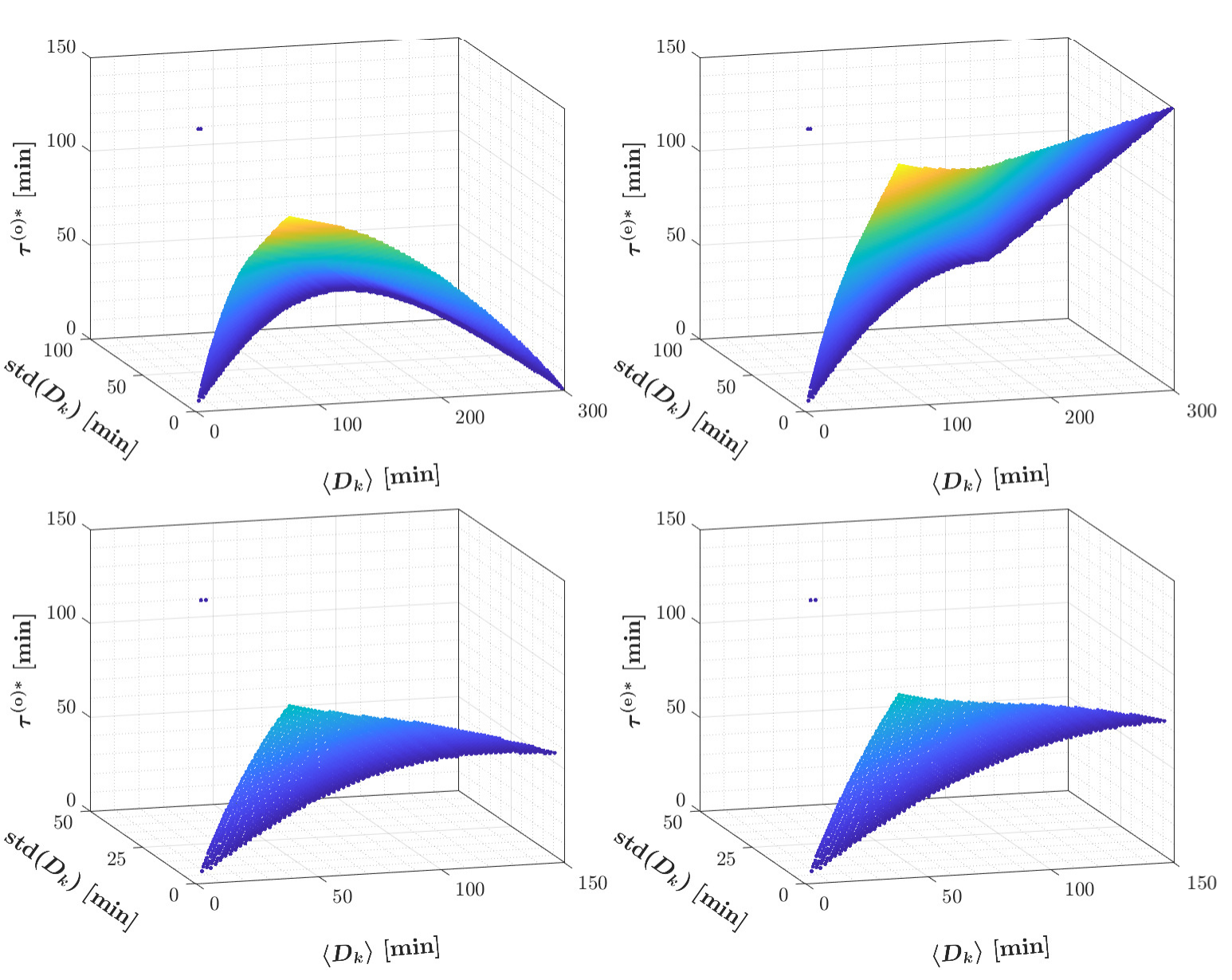}
\caption{Transition point $\tau^{\mathrm{(x)}*}$  versus the average congestion duration $\langle D_k \rangle$ and the standard deviation  $\mathrm{std}(D_k)$ of congestion durations for a length $T=300$ min of time series, where x=o for the left column and x=e for the right column. Each point in the figure originates from a $K\times T$ data matrix $I$, which is simulated with a given $D_\mathrm{max}$ and a given $D_\mathrm{min}$. For different points, $D_\mathrm{min}$ runs from 2 to $L-2$ and $D_\mathrm{max}$ runs from $D_\mathrm{min}+2$ to $L$, where $L=T=300$ min for top row and $L=T/2=150$ min for bottom row. The color visualizes the magnitude of $\mathrm{std}(D_k)$ with yellow for the largest value and dark blue for the smallest value.}
\label{fig6}
\end{center}
\end{figure}

\begin{figure}[htbp]
\centering
\includegraphics[width=\textwidth]{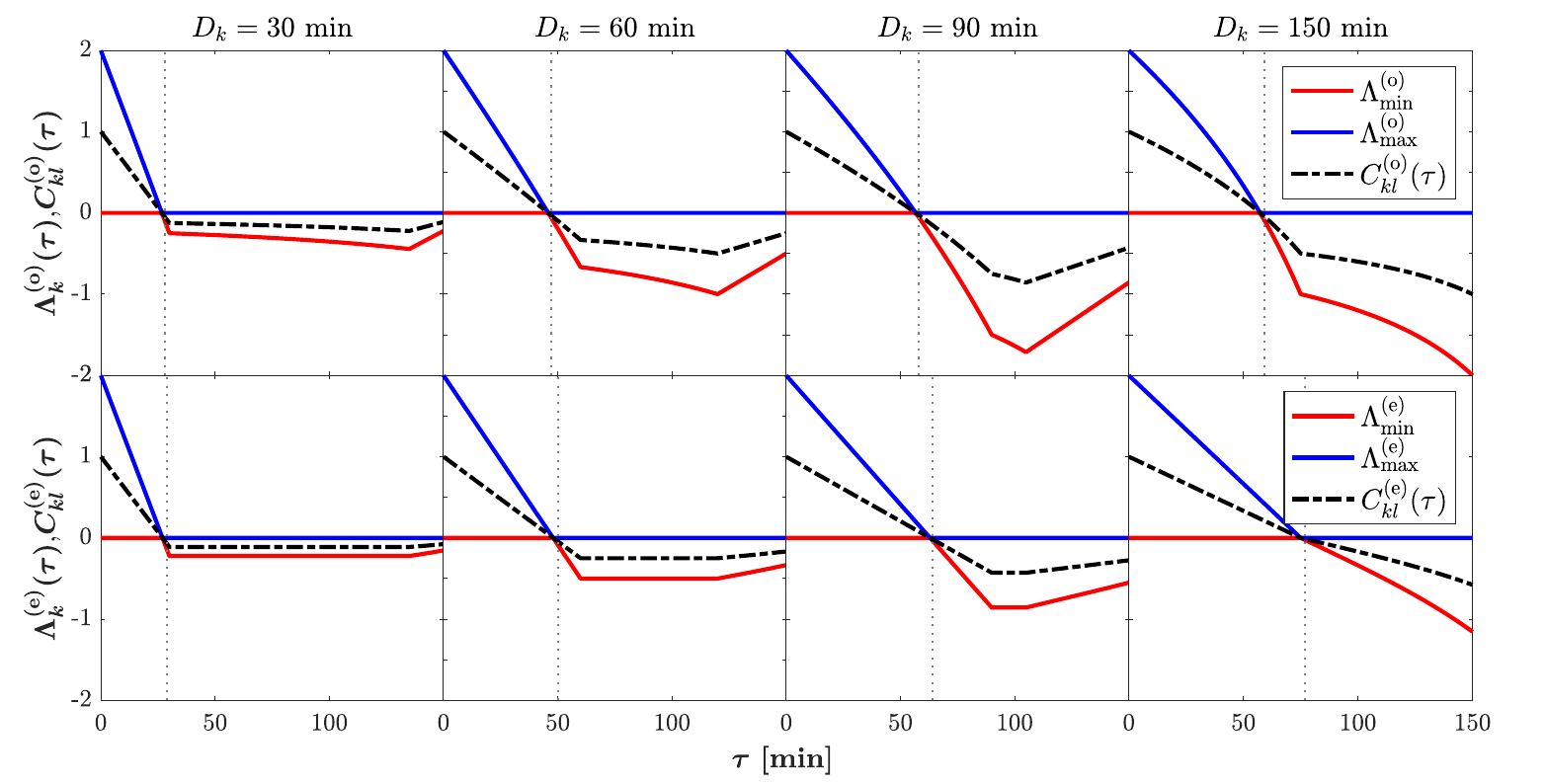}
\caption{Eigenvalues of $2\times 2$ original correlation matrices $\bar{C}^\mathrm{(o)}(\tau)$ and of $2\times 2$ extended correlation matrices $\bar{C}^\mathrm{(e)}(\tau)$ as well as the correlations $C_{kl}^\mathrm{(o)} (\tau)$ and $C_{kl}^\mathrm{(e)} (\tau)$  between sections $k$ and $l$, where $T = 300$ min and $D_k=D_l$.}
\label{fig7}
\end{figure}

\subsection{Spectral transitions at the vanishing of correlations}
\label{sec32}

Each simulated data matrix shown in figure~\ref{fig5} contains $K$ time series of indicators, where $K>2$. To simplify the situation, here we reduce the number of time series to two, i.e. $K=2$, and simulate $2\times T$ data matrices of indicators with $T=300$ min. We assume that the time series in the first row of the $2\times T$ data matrix is from section $k$ and the time series in the second row is from section $l$. The congestion for each time series is located symmetrically around the center of the time series of length $T$ with zero values. The other values in each time series are one, indicating free phases. For equal congestion duration, i.e., $D_k=D_l$, the original and extended correlation can be considered as the autocorrelation of the same time series with lengths $T$ and $T+\tau$, respectively. Therefore, the four elements at a certain time lag are the same, i.e.,  $C_{kk}^\mathrm{(x)}(\tau)=C_{kl}^\mathrm{(x)}(\tau)=C_{lk}^\mathrm{(x)}(\tau)=C_{ll}^\mathrm{(x)}(\tau)$, where x=o or e. In this setting, the correlation averaged over four elements is equal to any element in the $2\times 2$ correlation matrix $C^\mathrm{(x)}(\tau)$. Due to the symmetry of this correlation matrix, we have $C^\mathrm{(x)}(\tau)=\bar{C}^\mathrm{(x)}(\tau)$. The resulting two eigenvalues for each $2\times 2$ correlation matrix are shown in figure~\ref{fig7}. For congestion duration $D_k=D_l\leq T/2$, the transitions in the largest and the smallest eigenvalues intersect with the point when $C_{kl}^\mathrm{(x)}(\tau)=0$. The vanishing of the correlation between the two sections, therefore, signals the transition point of the dominant eigenvalue of a $2\times 2$ correlation matrix,
\begin{equation}
\tau^\mathrm{(x)*}=\mathrm{arg~min} \left| C_{kl}^\mathrm{(x)}(\tau)\right|  \ , \qquad \mathrm{x}=\mathrm{e,o}\ . 
\label{eq3.2.1}
\end{equation}

\begin{figure}[t]
\centering
\includegraphics[width=\linewidth]{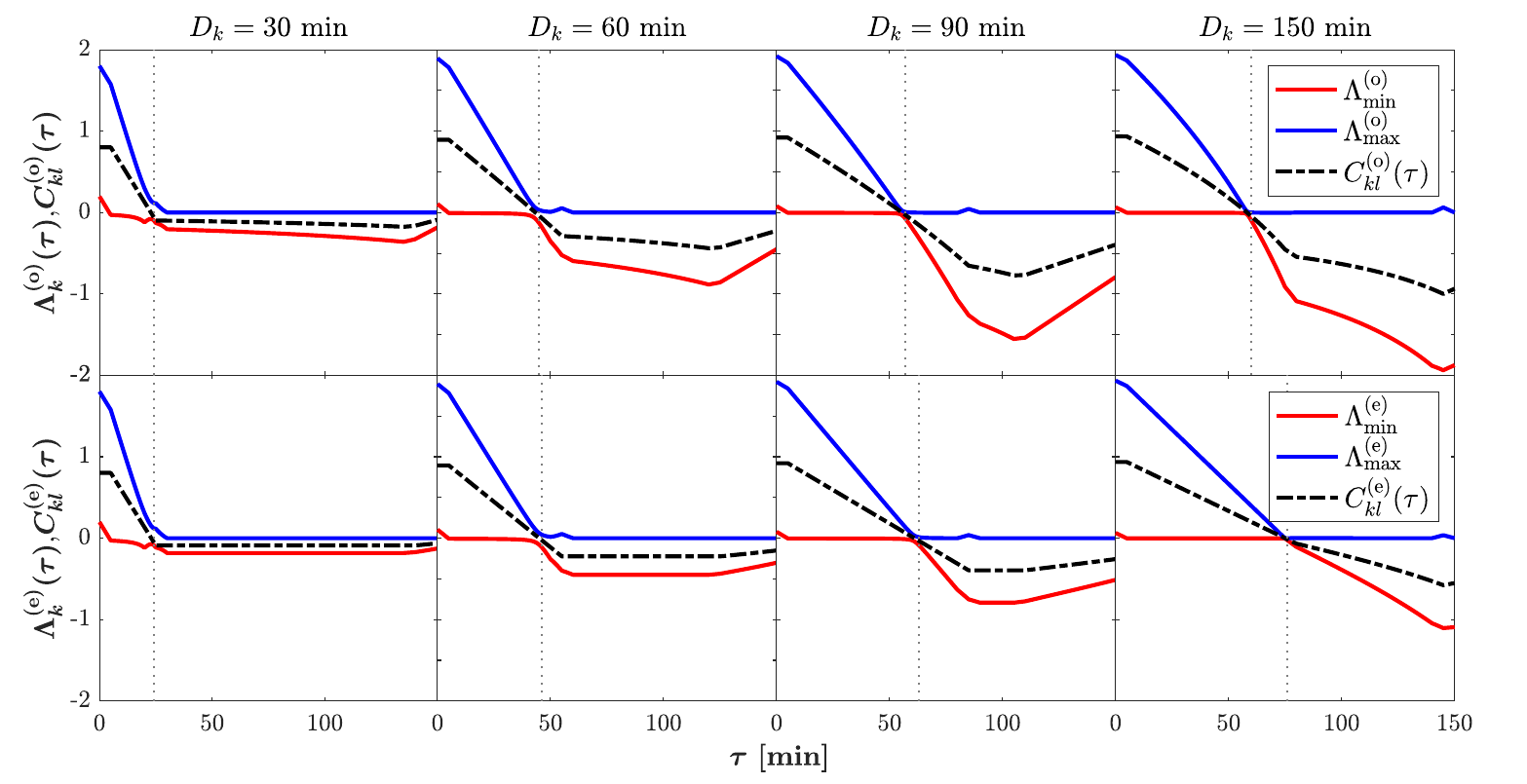}
\caption{Eigenvalues of $2\times 2$ original correlation matrices $\bar{C}^\mathrm{(o)}(\tau)$ and of $2\times 2$ extended correlation matrices $\bar{C}^\mathrm{(e)}(\tau)$ as well as the correlations $C_{kl}^\mathrm{(o)} (\tau)$ and $C_{kl}^\mathrm{(e)} (\tau)$  between sections $k$ and $l$, where $T = 300$ min and $D_l=D_k-10$ min.}
\label{fig8}
\end{figure}

\begin{figure}[!h]
\centering
\includegraphics[width=\linewidth]{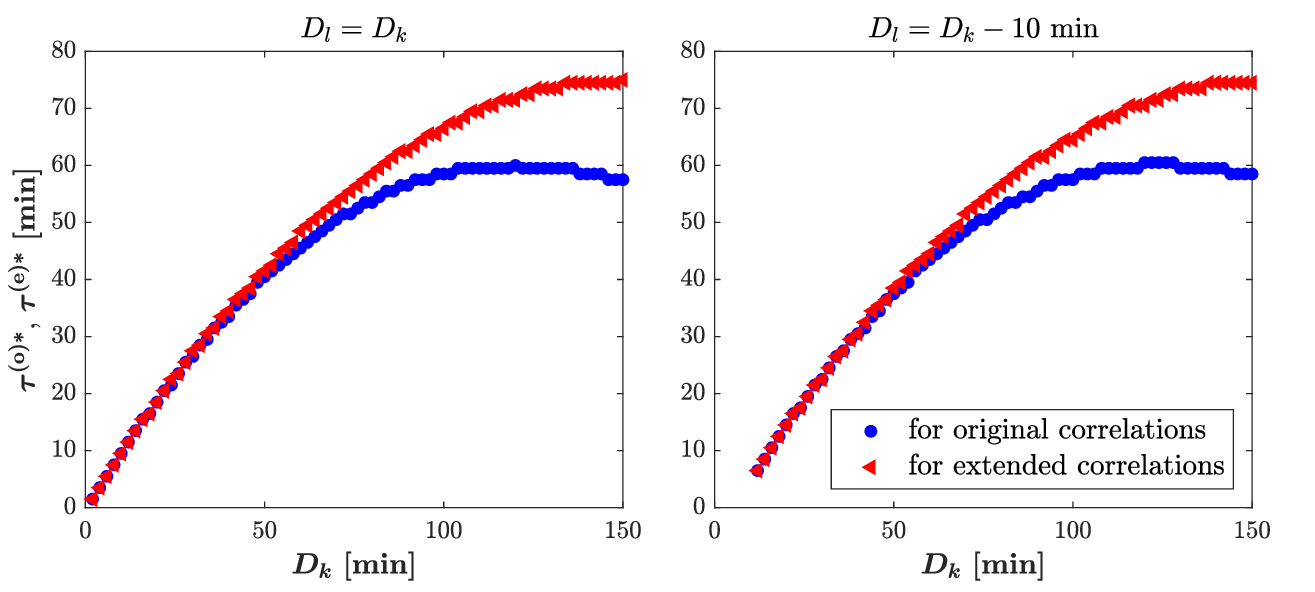}
\caption{The transition points $\tau^\mathrm{(o)*}$ and $\tau^\mathrm{(e)*}$ versus the congestion duration $D_k$ with respect to the total length of time series $T=300$ min, where the congestion durations of any two sections $k$ and $l$ are equal, i.e. $D_k=D_l$, in the left subfigure, and are unequal, i.e. $D_k=D_l+10$ min, in the right subfigure.}
\label{fig9}
\end{figure}

In reality, since the propagation and relief of congestion take time, the congestion durations for two sections are seldom the same, as shown in figure~\ref{fig4}. Here we set the propagation time to be $\Delta D/2$, such that $D_l=D_k-\Delta D$. The congestion for each section also lies symmetrically around the center of the time series. In this scenario, the four elements in each correlation matrix are different. We symmetrize each correlation matrix before obtaining the eigenvalues from $\bar{C}^\mathrm{(x)}(\tau)$. Setting $\Delta D/2=1$ min, we find the transition point in eigenvalues overlaps quite well with the point when $C_{kl}^\mathrm{(x)}(\tau)=0$, despite the fact that $C_{kl}^\mathrm{(x)}(\tau)$ is one of the off-diagonal elements in the $2\times 2$ correlation matrix instead of the average of four elements in this matrix. To enlarge the effect of propagation time on the transition points of eigenvalues, we consider $\Delta D=10$ min.  As shown in figure~\ref{fig8}, the transitions in the largest and the smallest eigenvalues are still very close to the point when $C_{kl}^\mathrm{(x)}(\tau)=0$. In contrast to the congestion duration $D_k$ with different values, $D_k=30,~60,~90$, and 150 min, the effect of duration difference $\Delta D$ is negligible. This shows that as long as the difference between two durations is short enough, e.g. $\Delta D\leq10$ min, the transition points in dominant eigenvalues can be estimated by equation~\eqref{eq3.2.1}.

Using the vanishing of correlations, we find a relation between the transition point $\tau^\mathrm{(o)*}$ or $\tau^\mathrm{(e)*}$ in the dominant eigenvalues and the congestion duration $D_k$ when $D_k\leq T/2$ and $T=300$ min, displayed in figure~\ref{fig9}. With increase of the congestion duration, the transition moves towards large values in a nonlinear fashion. For small congestion durations, the transition points for the original and extended correlations are close. For large congestion durations, a large deviation in the transition points emerges for the two kinds of correlations. The duration difference $\Delta D$ has little influence.

\section{Analytical treatment of the model for $2\times 2$ correlation matrices}
\label{sec4}

In our previous analyses, we explored the transition points of eigenvalues. We now consider a simplified, schematic version of the model that captures the features found. We normalize the indicators for original and extended correlations in section~\ref{sec41}. We model five scenarios with equal and unequal durations of congestion in sections~\ref{sec42} and \ref{sec43}, respectively. 

\subsection{Normalized indicators for original and extended correlations}
\label{sec41}

We define a relative congestion duration by normalizing to the length of time series 
\begin{equation}
\alpha_k=\frac{D_k}{T} \quad (0<\alpha_k\leq 1) 
\end{equation}
and accordingly relative time lags
\begin{equation}
\beta=\frac{\tau}{T} \quad (0<\beta\leq 1) \ . 
\end{equation}
For original correlations, the mean value and the standard deviation are calculated for the used time series of length $T-\tau$, not for the whole time series of length $T$. We express them in terms of $\beta$ and $\alpha_k$ as
\begin{equation}
\mu_{k}^\mathrm{(o)}=\frac{1}{T-\tau} \sum_{t=1}^{T-\tau} I_{k}(t) =\frac{T-\tau-D_k}{T-\tau}=1-\frac{\alpha_k}{1-\beta}
\label{eq4.2.1.1}
\end{equation}
and 
\begin{equation}
\sigma_{k}^\mathrm{(o)} =\sqrt{\frac{1}{T-\tau} \sum_{t=1}^{T-\tau}\Big(I_{k}(t)-\mu_{k}^\mathrm{(o)}\Big)^{2}}=\frac{\sqrt{(T-\tau-D_k)D_k}}{T-\tau}=\frac{\sqrt{(1-\beta-\alpha_k)\alpha_k}}{1-\beta} \ ,
\label{eq4.2.1.2}
\end{equation}
respectively. In an indicator time series $k$, $I_k^{\mathrm{(free)}} (t) =1$ for free traffic phases and $I_k^{\mathrm{(cong)}} (t) =0$ for congested traffic phases. Here and in the following calculations, we split the sum running from $t=1$ to $t=T-\tau$ or $T$ into three sums, running from 1 to $t_1$, $t_1+1$ to $t_2$, and $t_2+1$ to $T-\tau$ or $T$, in which $I_k(t)$ takes the values $I_k^{\mathrm{(free)}} (t)$, $I_k^{\mathrm{(cong)}} (t)$ and $I_k^{\mathrm{(free)}} (t)$, respectively. 
 For extended correlations, the time series of length $T$ is used. Thus, the mean value and the standard deviation, given by
\begin{equation} 
\mu_{k}^\mathrm{(e)} =\frac{1}{T} \sum_{t=1}^{T} I_{k}(t) =1-\frac{D_k}{T}=1-\alpha_k \ ,
\label{eq4.2.1.1}
\end{equation}
and 
\begin{equation}
\sigma_{k}^\mathrm{(e)} =\sqrt{\frac{1}{T} \sum_{t=1}^{T}\Big(I_{k}(t)-\mu_{k}^\mathrm{(e)} \Big)^{2}}=\sqrt{\left(1-\frac{D_k}{T}\right)\frac{D_k}{T}}=\sqrt{(1-\alpha_k)\alpha_k} \ ,
\label{eq4.2.1.2}
\end{equation}
are different from the ones for the original correlations. As seen in figure~\ref{fig10}, for each kind of correlation, given a non-zero time lag, the time series of section $l$ lags behind the time series of section $k$, but these two time series have the same mean values and standard deviations. We then normalize the indicators of free and congested phases for original and extended correlations to zero mean and unit standard deviation, 
\begin{eqnarray}
M_{k}^\mathrm{(o,free)}(t)&=&\frac{I_{k}^\mathrm{(free)}(t)-\mu_{k}^\mathrm{(o)}}{\sigma_{k}^\mathrm{(o)} }=\sqrt{\frac{\alpha_k}{1-\beta-\alpha_k}} \ ,
\label{eq4.2.1.4}
\\
M_{k}^\mathrm{(o,cong)}(t)&=&\frac{I_{k}^\mathrm{(cong)}(t)-\mu_{k}^\mathrm{(o)}}{\sigma_{k}^\mathrm{(o)}}=-\sqrt{\frac{1-\beta-\alpha_k}{\alpha_k}} \ ,
\label{eq4.2.1.5}
\\
M_{k}^\mathrm{(e,free)}(t)&=&\frac{I_{k}^\mathrm{(free)}(t)-\mu_{k}^\mathrm{(e)}}{\sigma_{k}^\mathrm{(e)} }=\sqrt{\frac{\alpha_k}{1-\alpha_k}} \  ,
\label{eq4.2.1.4}
\\
M_{k}^\mathrm{(e,cong)}(t)&=&\frac{I_{k}^\mathrm{(cong)}(t)-\mu_{k}^\mathrm{(e)}}{\sigma_{k}^\mathrm{(e)}}=-\sqrt{\frac{1-\alpha_k}{\alpha_k}} \ .
\label{eq4.2.1.5}
\end{eqnarray}
We notice that for original as well as for extended correlations, the normalized indicator of free traffic phases is the negative of the reciprocal of the normalized indicator of congested traffic phases, leading to  
$M_{k}^\mathrm{(o,free)}(t)M_{k}^\mathrm{(o,cong)}(t)=-1$ and $M_{k}^\mathrm{(e,free)}(t)M_{k}^\mathrm{(e,cong)}(t)=-1$. Furthermore, reducing $\tau$ to zero results in $\beta=0$ and yields $M_{k}^\mathrm{(o,free)}(t)=M_{k}^\mathrm{(e,free)}(t)$ and $M_{k}^\mathrm{(o,cong)}(t)=M_{k}^\mathrm{(e,cong)}(t)$.

\begin{figure}[tb]
\raggedright
Scenario 1\\
\hspace*{1cm}
\includegraphics[width=0.845\linewidth,left]{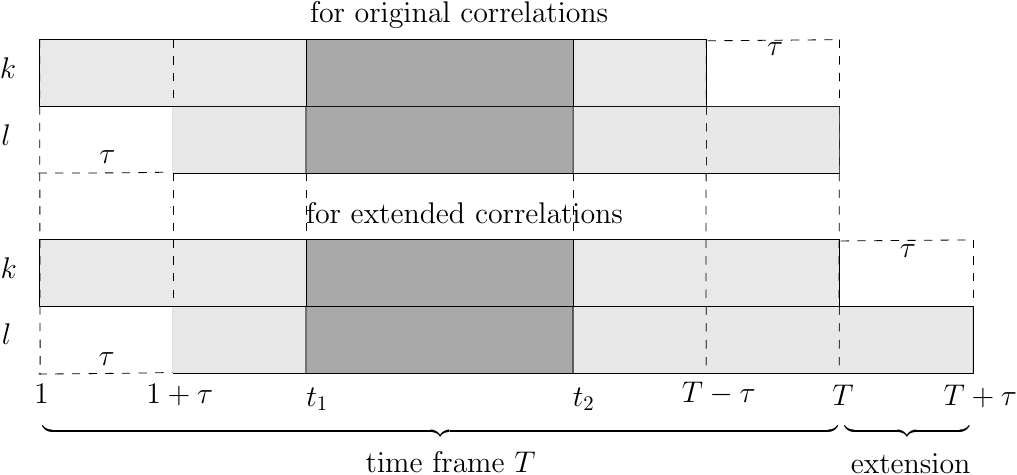}\\
\raggedright
Scenario 2\\
\hspace*{1cm}
\includegraphics[width=0.9\linewidth,left]{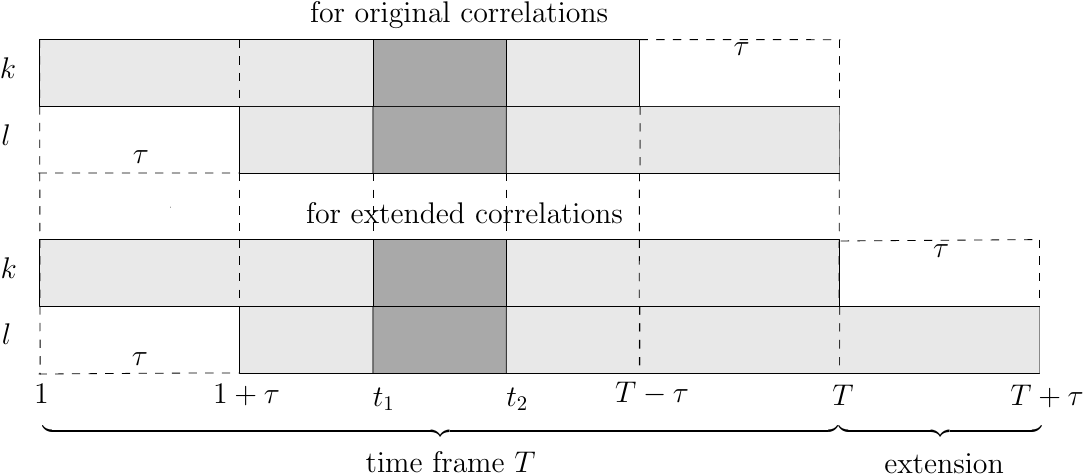}
\caption{Two indicator time series with equal duration of congestion for the calculation of original and extended correlations with time lag $\tau$. Light and dark grey regions represent the values for free and congested phases, respectively.}
\label{fig10}
\end{figure} 

\subsection{Scenarios with equal durations of congestion}
\label{sec42}

\begin{figure}[tb]
\centering
\includegraphics[width=1\linewidth]{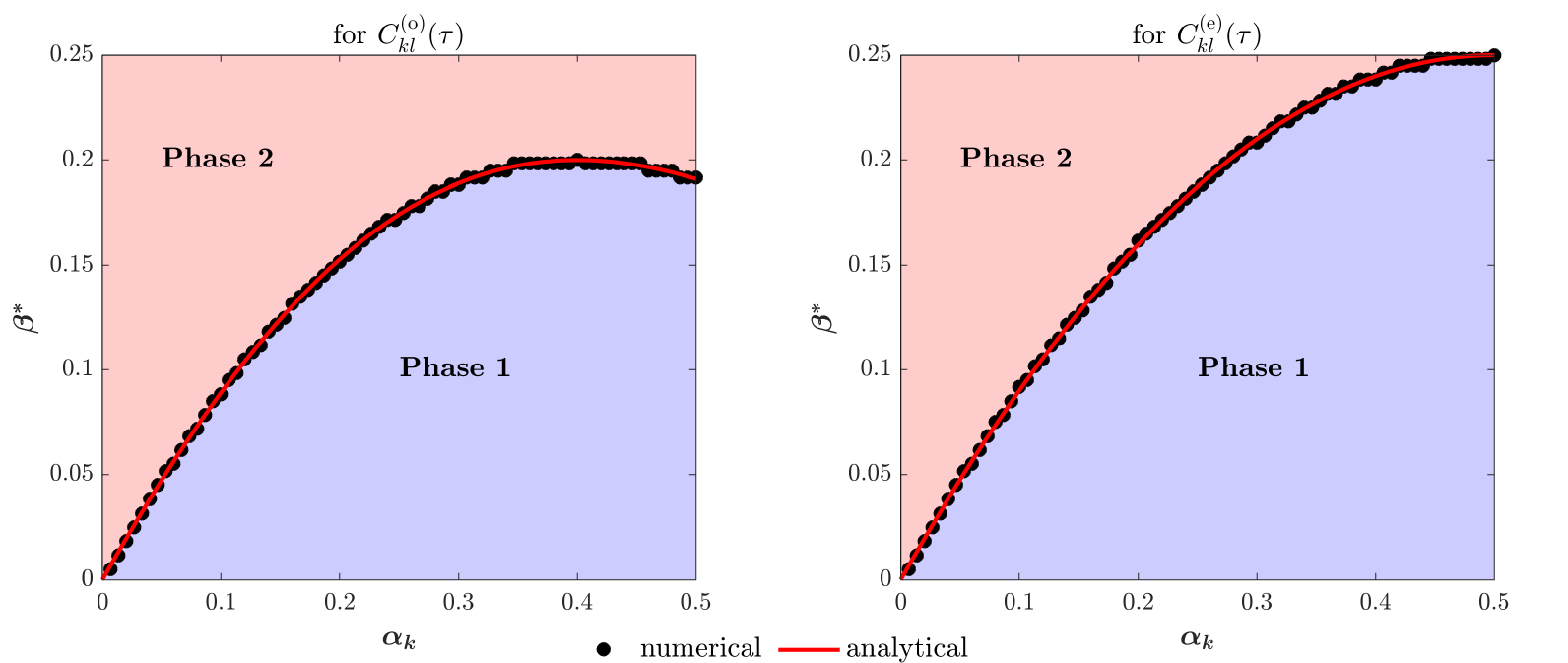}
\caption{Relative time lag $\beta^*$ of transition point versus the relative congestion duration $\alpha_k$ for scenario 1 with equal congestion duration. The different colors in the two plots indicate different spectral phases. Left: original correlations, right: extended correlations.}
\label{fig11}
\end{figure}

If the times at which the congestions on two sections $k$ and $l$ start and end coincide, e.g. both start at 16:00 and end at 17:00 on a workday, the two time series of indicators have equal congestion duration $D_k=D_l=t_2-t_1$. We refer to the scenario that the duration is larger than the positive time lag $\tau$, i.e. $D_k=D_l>\tau>0$, as  scenario 1 and to the scenario that the duration is smaller than or equal to $\tau$, i.e. $D_k=D_l\leq \tau$, as scenario 2, as sketched in figure~\ref{fig10}. For both scenarios, the time lag $\tau$ is always smaller than the start time $t_1$ of the congestion and the congestion is always symmetrically located in a time series with length $T$, i.e., $\tau<t_1$ and $t_1=T-t_2$. Here, $\tau<t_1$ also requires $\tau<T/2$. Otherwise, $D_k$ will be zero. 

According to the definitions of original correlations~\eqref{eq2.1.6} and extended correlations~\eqref{eq2.1.9}, we expand the sum of products of two normalized indicators into several parts, which are for free phases with free phases, congested phases with congested phases, free phases with congested phases and congested phases with free phases during different correlated times. In this way, we obtain expressions of correlations in terms of $\alpha_k$ and $\beta$. For scenario 1, the two kinds of correlations read 
\begin{eqnarray}\nonumber
C_{k l}^\mathrm{(o)}(\tau)&=& \frac{1}{T-\tau}\left(\sum_{t=1}^{t_{1}-\tau} M_{k}^\mathrm{(o,free )}(t) M_{l}^\mathrm{(o,free)}(t+\tau)+\sum_{t=t_{1}-\tau+1}^{t_{1}} M_{k}^\mathrm{(o,free)}(t) M_{l}^\mathrm{(o,cong)}(t+\tau)\right. \\ \nonumber
&+&\sum_{t=t_{1}+1}^{t_{2}-\tau} M_{k}^\mathrm{(o,cong)}(t) M_{l}^\mathrm{(o,cong)}(t+\tau)+\sum_{t=t_{2}-\tau+1}^{t_{2}} M_{k}^\mathrm{(o,cong)}(t) M_{l}^\mathrm{(o,free)}(t+\tau) \\
&+&\left.\sum_{t=t_{2}+1}^{T-\tau} M_{k}^\mathrm{(o,free)}(t) M_{l}^\mathrm{(o,free)}(t+\tau)\right) 
= \frac{-\alpha_k^2+(\alpha_k-\beta)(1-\beta)}{\alpha_k(1-\alpha_k-\beta)} \ ,
\label{eq4.2.1}
\end{eqnarray}
and 
\begin{eqnarray}\nonumber
C_{k l}^\mathrm{(e)}(\tau)&=& \frac{1}{T}\left(\sum_{t=1}^{t_{1}-\tau} M_{k}^\mathrm{(e,free )}(t) M_{l}^\mathrm{(e,free)}(t+\tau)+\sum_{t=t_{1}-\tau+1}^{t_{1}} M_{k}^\mathrm{(e,free)}(t) M_{l}^\mathrm{(e,cong)}(t+\tau)\right. \\ \nonumber
&+&\sum_{t=t_{1}+1}^{t_{2}-\tau} M_{k}^\mathrm{(e,cong)}(t) M_{l}^\mathrm{(e,cong)}(t+\tau)+\sum_{t=t_{2}-\tau+1}^{t_{2}} M_{k}^\mathrm{(e,cong)}(t) M_{l}^\mathrm{(e,free)}(t+\tau) \\
&+&\left.\sum_{t=t_{2}+1}^{T} M_{k}^\mathrm{(e,free)}(t) M_{l}^\mathrm{(e,free)}(t+\tau)\right) 
= \frac{\alpha_k-\alpha_k^2-\beta}{\alpha_k(1-\alpha_k)} \ ,
\label{eq4.2.2}
\end{eqnarray}
respectively. As seen in figure~\ref{fig7}, when the correlations vanish $C_{k l}^\mathrm{(o)}(\tau)=0$ and $C_{k l}^\mathrm{(e)}(\tau)=0$, the spectrum transition occurs. We therefore associate the relative time lag $\beta^*$ of transition point with the relative congestion duration $\alpha_k$, i.e., 
\begin{equation}
\beta^*=\frac{1}{2}\left(1+\alpha_k-\sqrt{(1-\alpha_k)^2+4\alpha_k^2} \right)
\label{eq4.2.3}
\end{equation}
for original correlations and
\begin{equation}
\beta^*=\alpha_k-\alpha_k^2
\label{eq4.2.4}
\end{equation}
for extended correlations. Figure~\ref{fig11} displays the dependence of relative time lag $\beta^*$ on relative congestion duration $\alpha_k$ simulated by equations~\eqref{eq4.2.3} and \eqref{eq4.2.4}, Here we restrict the congestion duration to half the length of the time series, i.e., $\alpha_k\leq 0.5$, as analyzed at the end of section~\ref{sec31}. For both kinds of correlations, the analytical results perfectly match with the numerical ones. The dependent curve is a transition curve that is connected by all transition points and separates the system into two spectral phases. Below the curve, it is the first spectral phase that the largest eigenvalue dominates in the system. Above the curve, it transforms to the second spectral phase that the smallest eigenvalue dominates. According to equations~\eqref{eq4.2.3} and \eqref{eq4.2.4},  the maximal $\beta^*=0.2$ at $\alpha_k=0.4$ for original correlations and the maximal $\beta^*=0.25$ at $\alpha_k=0.5$ for extended correlations, which means the maximal transition points are at $\tau=60$ min and at $\tau=75$ min for original and extended correlation, respectively. The two values are close to the simulated transition points $\tau^\mathrm{(o)*}=59.5$ min and $\tau^\mathrm{(e)*}=69.5$ min for scenario (c) with $D_\mathrm{max}=T/2$ min and $D_\mathrm{min}\approx T/4$ min listed in table~\ref{tab1}. The slight differences in transition points are account of different constraints, as the simulation in scenario (c) is for 39 sections with gradually changed congestion durations, while the model in scenario 1 only for two sections with the same congestion duration. Despite of this, we notice that the congestion duration $T/2$ plays a critical role in the transition of eigenvalues.

For scenario 2, despite the fact that the correlations in terms of $\beta^*$ and $\alpha_k$ can be derived, we cannot use the vanishing of correlations to find a relation between $\beta^*$ and $\alpha_k$ under the constraints $0<\alpha_k\leq 0.5$ and $0<\beta<0.5$.

\subsection{Scenarios with differing durations of congestion}
\label{sec43}

\begin{figure}[htbp]
\raggedright
Scenario 3 \\
\hspace*{1cm}
\includegraphics[width=0.845\linewidth,left]{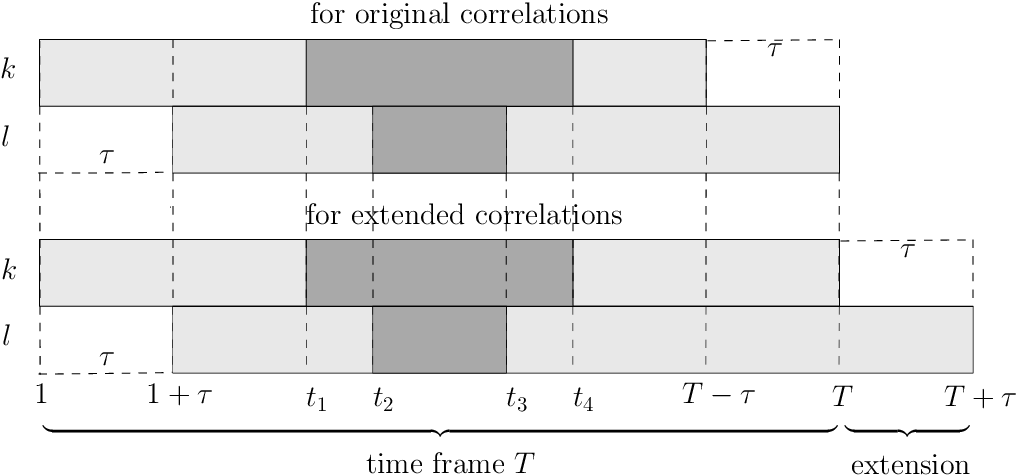}\\
\raggedright
Scenario 4\\
\hspace*{1cm}
\includegraphics[width=0.779\linewidth,left]{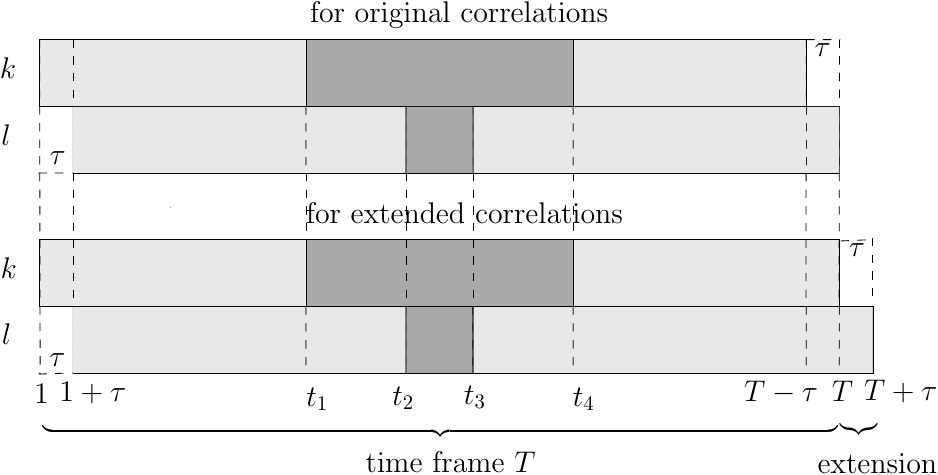}\\
\raggedright
Scenario 5\\
\hspace*{1cm}
\includegraphics[width=0.9\linewidth,left]{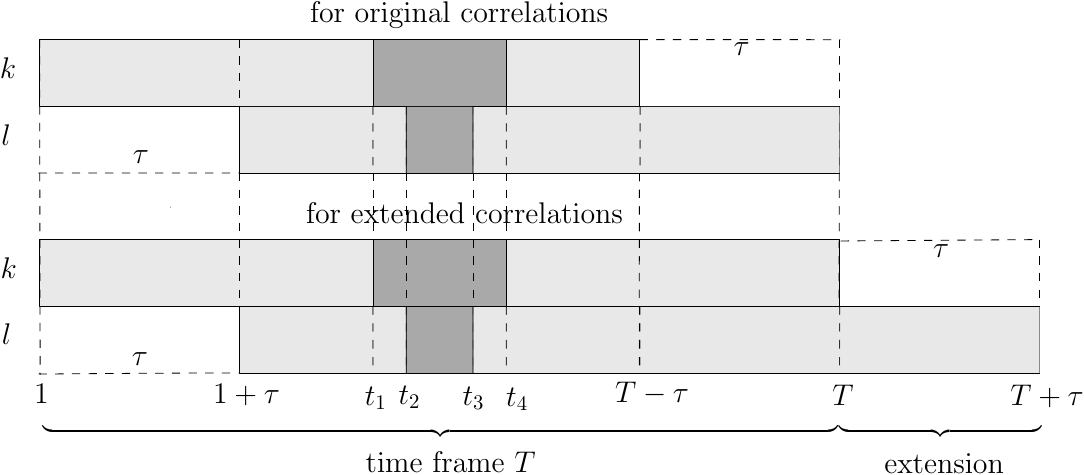}
\caption{Two indicator time series with unequal duration of congestion for the calculation of original and extended correlations with time lag $\tau$. Light and dark grey regions represent the values for free and congested phases, respectively. }
\label{fig12}
\end{figure}

If the congestion on section $k$ starts first and propagates to section $l$,  the congestion on section $l$ will start a little later but will end earlier than on section $k$, leading to $D_k>D_l$, where $D_k=t_4-t_1$ and $D_l=t_3-t_2$, as shown in figure~\ref{fig12}. For instance, the congestion duration for section $k$ is from 15:50 to 17:10 and for section $l$ from 16:00 to 17:00. Here we assume the congestion is symmetrically located in each time series of length $T$, such that $t_1=T-t_4$, $t_2=T-t_3$ and $t_2-t_1=t_4-t_3=(D_k-D_l)/2$. In the following, we consider three scenarios, namely scenarios 3, 4 and 5, where in scenarios 3 and 4, the time lag is smaller than or equal to the short congestion duration $0<\tau\leq D_l<D_k$, but in scenario 5, the time lag larger than or equal to the long congestion duration $0<D_l<D_k\leq \tau$. Moreover, in scenarios 3 and 5, the time lag is larger than or equal to the half difference of congestion duration $(D_k-D_l)/2\leq \tau$, but in scenario 4, it is smaller than half the difference of congestion duration $(D_k-D_l)/2>\tau$. To distinguish the five scenarios clearly, we list the constraints for each of them in table~\ref{tab2}. In terms of the relative congestion durations $\alpha_k$ and $\alpha_l$ for sections $k$ and $l$ as well as the relative time lag $\beta$, we work out the correlation coefficient,
\begin{eqnarray}\nonumber
C_{k l}^\mathrm{(o)}(\tau)&=& \frac{1}{T-\tau}\left(\sum_{t=1}^{t_{2}-\tau} M_{k}^\mathrm{(o,free )}(t) M_{l}^\mathrm{(o,free)}(t+\tau)+\sum_{t=t_2-\tau+1}^{t_1} M_{k}^\mathrm{(o,free)}(t) M_{l}^\mathrm{(o,cong)}(t+\tau)\right. \\ \nonumber
&+&\sum_{t=t_1+1}^{t_3-\tau} M_{k}^\mathrm{(o,cong)}(t) M_{l}^\mathrm{(o,cong)}(t+\tau)+\sum_{t=t_3-\tau+1}^{t_4} M_{k}^\mathrm{(o,cong)}(t) M_{l}^\mathrm{(o,free)}(t+\tau) \\
&+&\left.\sum_{t=t_4+1}^{T-\tau} M_{k}^\mathrm{(o,free)}(t) M_{l}^\mathrm{(o,free)}(t+\tau)\right) 
= \frac{(\alpha_l-2\beta)(1-\beta)+\alpha_k(1-2\alpha_l-\beta)}{2\sqrt{\alpha_k(1-\alpha_k-\beta)}\sqrt{\alpha_l(1-\alpha_l-\beta)}}	
\label{eq4.3.1}
\end{eqnarray}

\begin{table}[t!]
\caption{Five scenarios with their constraints}
\begin{center}
\begin{tabular*}{0.91\textwidth}{|c|@{\hskip 0.5cm} l@{\hskip 0.5cm}| @{\hskip 0.5cm} l@{\hskip 0.5cm}| @{\hskip 0.5cm}l@{\hskip 0.5cm}|}
\hline
scenario & constraint 1 & constraint 2 & constraint 3 \\
\hline
Scenario 1&	$0<\tau<D_k=D_l$ & $D_k=D_l=t_2-t_1$ & $\tau<t_{1}=T -t_2$	\\
\hline	
Scenario 2 &	$0<D_k=D_l\leq \tau<T/2$  & $D_k=D_l=t_2-t_1$  & $\tau<t_1=T -t_2$	 \\ 
\hline
\multirow{2}{*}{Scenario 3} &	\multirow{2}{*}{$0<\tau\leq D_l<D_k$}     & $(D_k-D_l)/2=t_2-t_1$  & $\tau<t_1=T -t_4$   \\
 && $=t_4-t_3 \leq \tau$ & $<t_2=T-t_3$ \\
 \hline
 \multirow{2}{*}{Scenario 4} &	\multirow{2}{*}{$0<\tau\leq D_l<D_k$}  &  $(D_k-D_l)/2=t_2-t_1$ & $\tau<t_1=T -t_4$  \\
 & & $=t_4-t_3 > \tau$ &  $<t_2=T-t_3$ \\
 \hline
\multirow{2}{*}{Scenario 5} &	\multirow{2}{*}{$0<D_l<D_k\leq\tau<T/2$}  &  $(D_k-D_l)/2=t_2-t_1$ & $\tau<t_1=T -t_4$   \\
 &  & $=t_4-t_3 \leq \tau$ & $<t_2=T-t_3$\\
\hline
\end{tabular*}
\end{center}
\label{tab2}
\end{table}%

\begin{figure}[t!]
\includegraphics[width=\linewidth]{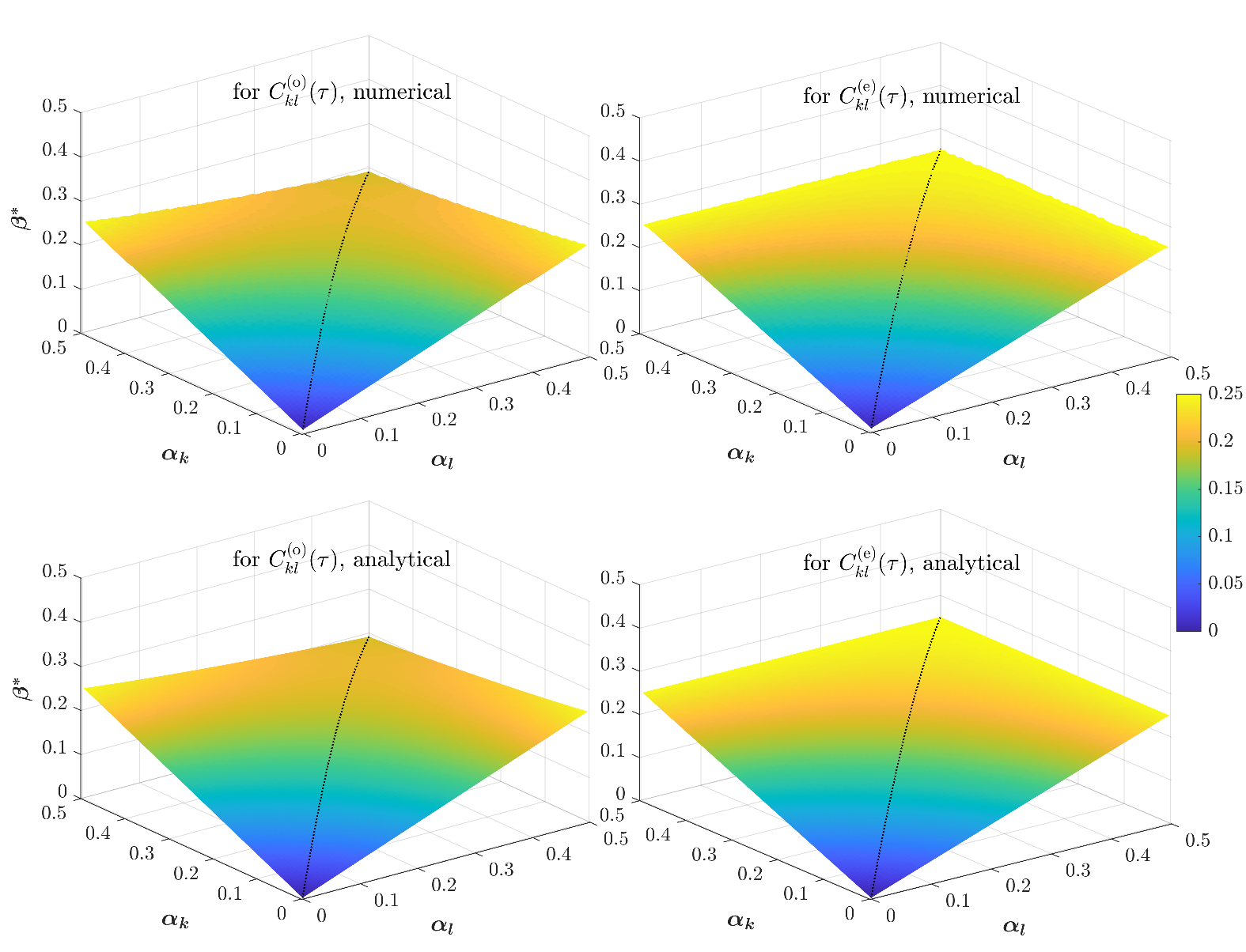}
\caption{Relative time lag $\beta^*$  versus relative congestion duration $\alpha_k$ and $\alpha_l$ for scenario 3 with unequal congestion duration. The dotted lines correspond to scenario 1 when $\alpha_k=\alpha_l$. Left: original correlations, right: extended correlations, top: numerical results, bottom: analytical results.}
\label{fig13}
\end{figure}

for original correlations, and 
\begin{eqnarray}\nonumber
C_{k l}^\mathrm{(e)}(\tau)&=& \frac{1}{T}\left(\sum_{t=1}^{t_{2}-\tau} M_{k}^\mathrm{(e,free )}(t) M_{l}^\mathrm{(e,free)}(t+\tau)+\sum_{t=t_2-\tau+1}^{t_1} M_{k}^\mathrm{(e,free)}(t) M_{l}^\mathrm{(e,cong)}(t+\tau)\right. \\ \nonumber
&+&\sum_{t=t_1+1}^{t_3-\tau} M_{k}^\mathrm{(e,cong)}(t) M_{l}^\mathrm{(e,cong)}(t+\tau)+\sum_{t=t_3-\tau+1}^{t_4} M_{k}^\mathrm{(e,cong)}(t) M_{l}^\mathrm{(e,free)}(t+\tau) \\
&+&\left.\sum_{t=t_4+1}^{T} M_{k}^\mathrm{(e,free)}(t) M_{l}^\mathrm{(e,free)}(t+\tau)\right) 
= \frac{\alpha_k+\alpha_l-2\alpha_k\alpha_l-2\beta}{2\sqrt{\alpha_k(1-\alpha_k)}\sqrt{\alpha_l(1-\alpha_l)}}	
\label{eq4.3.2}
\end{eqnarray}
for extended ones. To find the transition point, we use the vanishing of the original correlations, i.e. $C_{k l}^\mathrm{(o)}(\tau)=0$, and thereby obtain
\begin{equation}
\beta^*=\frac{1}{4}\left(2+\alpha_k+\alpha_l-\sqrt{(\alpha_k+\alpha_l-2)^2+16\alpha_k\alpha_l} \right) \ .
\label{eq4.3.3}
\end{equation}
Analogously, we arrive at 
\begin{equation}
\beta^*=\frac{1}{2}\left(\alpha_k+\alpha_l-2\alpha_k\alpha_l \right)
\label{eq4.3.4}
\end{equation}
by solving $C_{k l}^\mathrm{(e)}(\tau)=0$ for extended correlations.

In the range $0<\alpha_k\leq 0.5$, $0<\alpha_l\leq 0.5$ and $0<\beta< 0.5$, the dependence of $\beta^*$ on $\alpha_k$ and $\alpha_l$ manifests itself as a surface shown in a three-dimensional plot in figure~\ref{fig13}. The analytical results are very close to the numerical ones. The colors of the surface correspond to values of $\beta^*$ regardless of how $\alpha_k$ and $\alpha_l$ combine. Equations~\eqref{eq4.3.3} and \eqref{eq4.3.4} with $\alpha_k=\alpha_l$ result in equations~\eqref{eq4.2.3} and \eqref{eq4.2.4}. The black dotted line on surfaces in figure~\ref{fig13} corresponds to scenario 1 with equal congestion duration. Each surface in figure~\ref{fig13} separates spectral phase 1 from spectral phase 2. The former, under the surface, is governed by the largest, positive eigenvalue while the latter, above the surface, is dominated by the smallest, negative eigenvalue. Obviously, when the congestion durations for either or both sections are around $T/2$, there are many high values of $\beta^*$ colored red on the surface, implying large transition points, before which the largest eigenvalue prevails over the smallest one for a long time. In contrast, when the congestion durations on both sections approach 0, some low values of $\beta^*$ colored blue on the surface are visible, suggesting small transition points, which further reveals the smallest eigenvalues are quickly victorious over the largest one. In other words, the largest eigenvalue dominates longer if the congestion duration of either section is around $T/2$ but shorter if the congestion durations of both sections approach 0. From another point of view, when given any two of the three quantities $\beta^*$, $\alpha_k$ and $\alpha_l$, we are able to infer the remaining quantity from the three-dimensional surface, especially the congestion duration $\alpha_k$ or $\alpha_l$. Notably, when $\alpha_l=\alpha_k$, equations~\eqref{eq4.3.3} and \eqref{eq4.3.4} are simplified to equations~\eqref{eq4.2.3} and \eqref{eq4.2.4}, respectively.

For scenarios 4 and 5, a relation that fulfills the constraints $0<\alpha_k\leq 0.5$, $0<\alpha_k\leq 0.5$ and $0<\beta< 0.5$ at the same time is absent among $\alpha_k$, $\alpha_l$ and $\beta^*$.

\section{Conclusions}
\label{sec5}

The congestion propagation to neighbouring motorway sections gives rise to the correlated traffic behavior. The full time-lagged correlation matrix of a motorway network encodes the congestion characteristics among other information. We showed how to reveal them by the interesting spectra. 

A correlation matrix without time lags is symmetric and semi-positive definite, resulting in real and positive eigenvalues. With a time lag, however, this correlation matrix becomes asymmetric and the resulting eigenvalues are complex. To obtain real eigenvalues, one can addresses the time-lagged correlation matrix either by singular value decomposition or by eigenvalue decomposition after the symmetrization of the correlation matrix. Here, we took up the latter approach for the time-lagged correlation matrix of velocities. In empirical studies, we found a transition of the dominant eigenvalue from a positive value to a negative one as function of the time lag. This transition is related to the average of all elements in the correlation matrix. We set up a model by using indicator time series of traffic phases. Our numerical simulation show that this transition occurs when the average correlation (or the cross-correlation) vanishes for the dimension of the correlation matrix larger than (or equal to) two. Moreover, our numerical simulations of scenarios with different congestion durations reveal a nonlinear dependence of the transition point on the congestion duration. The nonlinear dependence for the latter with $2\times 2$ time-lagged correlation can be analytically described by a simple, schematic indicator model that we set up. With this analytical model, the congestion duration can be readily extracted from spectral transition of the correlation matrices.

In carrying out these studies, we also studied a generic statistics issue occurring when looking at time-lagged correlations. When the time lag is very large comparing to the fixed length of time series, the statistics become rather poor. We put forward an extended time-lagged correlation, relative to the original, widely used one. In the extended time-lagged correlation we used the time series of the full length as the leading time series, and the one including additional data points at the end as the lagged time series, such that leading and lagged time series always have the same length. By comparing the two kinds of time-lagged correlations, we found the extended correlation is smaller than the original one at a given time lag. Furthermore, we found when the congestion duration is larger than the half time series, a large divergence emerges in the nonlinear dependence of the spectral transition point on the congestion duration. This phenomenon implies a limitation of congestion duration to half of the time series in the analysis of its relation with spectral transition.

\section*{Acknowledgments}

We gratefully acknowledge the support from Deutsche Forschungsgemeinschaft (DFG ) within the project ``Korrelationen und deren Dynamik in Autobahnnetzen'' (No. 418382724). We thank Strassen.NRW for providing the empirical traffic data.

\addcontentsline{toc}{section}{References}

\end{document}